\documentclass{jfm}
\usepackage{graphicx}
\usepackage{epstopdf, epsfig}
 \usepackage{amssymb}
\usepackage{comment}
\usepackage{amsmath}
\usepackage{subcaption}
\usepackage{xcolor}
 \usepackage{hyperref}

\newcommand{\bv}[1]{\boldsymbol{#1}}

\shorttitle{SSD for Turbulent Ekman Layer} 
\shortauthor{Eojin Kim and others} 

\title{Statistical State Dynamics Based Study of the Turbulent Ekman Layer}

\author
 {
 Eojin Kim\corresp{\email{ekim@g.harvard.edu}}\aff{1},
 Brian F. Farrell\aff{1}
  }

\affiliation
{
\aff{1}
Department of Earth and Planetary Sciences, Harvard University, Cambridge, MA~02138, USA

}
\begin{document}
\maketitle

\begin{abstract}

Streamwise roll and streak structures (RSS) are prominent features observed in both atmospheric and oceanic planetary boundary layers (PBL) as well as in laboratory-scale wall-bounded shear flows. Despite their structural similarity across these systems, the mechanisms responsible for forming and sustaining the RSS remain debated. This study demonstrates that the same turbulence-sustaining mechanism previously identified in wall-bounded shear flows using the Statistical State Dynamics (SSD) formulation of the Navier–Stokes equations \citep{Farrell-Ioannou-2012, Farrell-Ioannou-2016-bifur} also operates in the Ekman layer. By extending the SSD-based stability analysis methods previously used for studying roll formation in wall-bounded shear flows to the Ekman layer, we show that the well-known Reynolds-stress–driven instability mechanism in wall-bounded turbulence acts together with inflectional instability to produce and sustain RSS in the Ekman layer. These results enhance the mechanistic understanding of RSS formation and evolution in the turbulent Ekman layer and provide a fundamental link between geophysical Ekman-layer turbulence and turbulence in engineering-scale shear flows.

\end{abstract}

\section{Introduction}
Improving understanding of turbulence in the planetary boundary layer (PBL) is important because the exchange of momentum and tracers between the surface and the deeper atmosphere and ocean is governed by transport processes driven by PBL turbulence dynamics.  One phenomenon that relies on these transports is the dynamics of tropical cyclones (TCs), especially the occurrence of rapid intensification—a process whose underlying mechanisms remain theoretically challenging to explain. In addition, cyclones cause extensive damage through extreme surface winds, wind stress-driven storm surges, and flooding patterns that are strongly affected by the large-scale coherent velocity structures formed within the planetary boundary layer.
Beginning with the initial Doppler radar observations of Hurricane Fran \citep{Wurman1998}, it has gradually become more widely recognized that the tropical cyclone boundary layer (TCBL) contains a substantial component of coherent boundary layer structure consisting of roll streak
structures (RSS) aligned closely with the mean wind.  Doppler radar observations of Hurricane Fran’s rolls, shown in \citet{Wurman1998}, illustrate RSS in the tropical cyclone boundary layer, but with scales toward the smaller end of the observed distribution. These rolls have a horizontal wavelength of approximately $600\,\mathrm{m}$, a roll velocity of about $4\,\mathrm{m\,s^{-1}}$, and associated azimuthal wind speed variations across the rolls of roughly $30\,\mathrm{m\,s^{-1}}$. More typical RSS in the TCBL span a range of scales and intensities, with mean properties characterized by a horizontal wavelength of about $1.4\,\mathrm{km}$, a depth near $700\,\mathrm{m}$, a roll velocity of about $3\,\mathrm{m\,s^{-1}}$, and azimuthal wind speed variations across the rolls of around $14\,\mathrm{m\,s^{-1}}$ \citep{Morrison2005}.

While observations of the TCBL demonstrate the presence of RSS both in the inner and outer core regions, the related problem of RSS formation in the Ekman layer—where geostrophic balance replaces gradient-wind balance—offers an appealing initial framework for extending our understanding of RSS formation in the PBL. We anticipate that similar RSS formation mechanisms act in both environments, with the TCBL distinguished by the presence of a manifold of inertial instabilities in the inner-core region. The deepening process is affected by RSS-mediated transport both in the inner and in the outer core, where the inertial manifold remains stable. Motivated by the intrinsic importance and relative simplicity of the RSS formation and equilibration in the Ekman layer turbulence problem, we begin our study by focusing on the Coriolis-balanced Ekman boundary layer in which the radial wind profile contains an inflection point that supports an RSS instability \citep{Lilly1966, Brown1980}.
Although this problem has been extensively examined, attempts to explain observed RSS in the PBL solely through Ekman layer instability still encounter the difficulty noted early by \citet{Brown1970}: “...Perhaps the most serious setback to the successful ... analysis lies in the failure of the elementary equilibrium balance ... solved by Ekman (1905) to be verified observationally...”.
In Figure 1b of the same study, a sounding associated with roll formation, as indicated by cloud street observations, exhibits essentially neutral stability and minimal wind veer with height. The observed flow departs so strongly from the laminar Ekman profile that it casts doubt on the applicability of stability analyses based on that profile for explaining roll structures observed in the equilibrated turbulent PBL and it has long been recognized that identifying a roll-type instability in equations linearized about the laminar Ekman solution is, by itself, insufficient to account for the rolls that appear in the turbulent, equilibrated boundary layer. At the same time that analytical understanding of RSS dynamics remains incomplete, recent large eddy simulation (LES) studies \citep{Gao2014,Gao2015} demonstrate that simulated RSS dynamics in the TCBL yields substantial discrepancies with observations. These findings highlight the need to improve understanding of RSS dynamics in the PBL.

While both theory and simulation studies commonly attribute initial roll formation in neutrally stable stratification to an inflectional instability \citep{Gao2014},  these studies provide no guidance for connecting the finite amplitude nonlinear statistically equilibrated turbulent RSS state to the linear instabilities of the laminar Ekman profile, which itself is not observed.
A comprehensive understanding of RSS turbulence dynamics, including connecting the origin and equilibration processes, is essential to account for the impact of the RSS on the PBL.\\
Equilibrium mean shear flows in wall-bounded turbulence are similar to observed mean turbulent PBL shear flows in being typically linearly stable. Nevertheless, the streamwise RSS produces strong momentum transport and mixing across the layer similar to that found in wall-bounded shear flow turbulence \citep{Kline1962} \citep{Smith1983}.
Because the RSS is not an unstable mode of the linearized Navier–Stokes equations (NSE) for laminar shear flow, its ubiquity in wall-bounded turbulence was not analytically explained until the development of optimal perturbation analysis, which demonstrated that the RSS corresponds to the structure of optimal transient growth \citep{Butler-Farrell-1992, Reddy1993}.  Later, it was discovered that the RSS is unstable in turbulent shear flow, as became evident when the NS equations were expressed in the statistical state dynamics (SSD) framework. This unstable RSS emerges through a mechanism in which Reynolds stresses, associated with a weakly sustained, unstructured background turbulence field, are systematically organized by a perturbation RSS mode to drive roll structures that are correctly oriented to sustain the lift-up mechanism underlying the RSS modal instability in the SSD \citep{Farrell-Ioannou-2012,Farrell-Ioannou-2016-bifur}. These SSD modal instabilities with RSS structure were found to equilibrate into fixed-point, finite-amplitude RSS configurations or, depending on parameter values, to transition to chaotic attractor dynamics, including self-sustaining turbulent states \citep{Farrell-Ioannou-2012}.
\\
Building on the success of SSD-based analyses in characterizing the dynamics and maintenance of RSS in wall-bounded shear flows, SSD theory has more recently been extended to explore RSS dynamics in geophysical flows, including applications to Langmuir turbulence \citep{Kimlangmuir} and Eady frontal turbulence \citep{Kimeadypart1, Kimeadypart2}.  As noted above, conventional studies that trace the origin of RSS to linear instability have largely failed to explain how the statistically averaged turbulent state is established and sustained. One effort to fill this gap, by \citet{Foster2005}, employed a weakly nonlinear Landau analysis to obtain a nonlinearly equilibrated state; however, this equilibrium differs substantially from the observed turbulent mean state. An advantage of SSD theory is that it naturally provides a unified framework that incorporates both the traditional inflectional instability mechanism and the SSD instability mechanism, offering a synergistic description of these instabilities. Importantly, SSD intrinsically includes mechanisms responsible fo the nonlinear equilibration of these instabilities, thereby delivering a comprehensive SSD-based analysis of both the origin and the nonlinear statistical mean equilibrium state of the RSS in the PBL. The observed turbulent mean velocity profile in the PBL is a specific nonlinear equilibrium, it is one profile selected from an infinite family of candidate stable profiles. Because the SSD equations are nonlinear, they capture both the onset of instability and the particular equilibrated statistical steady state that emerges uniquely from this infinite set of possible stable states. Moreover, because the SSD equilibrium includes both the mean profile and its associated fluctuation covariances, it simultaneously determines both the profile and the fluxes that sustain it.

In this work, we extend SSD theory to examine RSS dynamics in the Ekman layer of the PBL, addressing the initial development of the RSS, the establishment of the equilibrated mean RSS, the conditions under which a self-sustaining turbulent PBL state can persist, the emergence of statistical mean fluctuation covariances, and the influence of these covariances on transport—particularly the transport of momentum and water vapor.



\section{Governing Equations for the PBL}

We examine the dynamics of a turbulent PBL using coordinates x, y, and z, representing the streamwise, cross-stream, and spanwise directions, respectively. The corresponding velocity components are denoted by u, v, and w, with unit vectors i, j, and k, respectively. The governing equations are:

\begin{equation}
\frac{\partial \mathbf{u}}{\partial t}+(\mathbf{u}\cdot \nabla)\mathbf{u}=-\nabla p -2 \Omega \hat{j} \times \mathbf{u}+ \nu \Delta \mathbf{u}
\end{equation}\\
where $\Delta:=(\partial_{xx}+\partial_{yy}+\partial_{zz})$, $\mathbf{u}=u\hat{i}+v\hat{j}+w\hat{k}$,  $\nu$ is kinematic viscosity and $\Omega$ represents planetary rotation rate at the given latitude.\\

Geostrophic balance is recovered at the top of the PBL
\begin{equation}
2 \Omega \hat{y} \times \mathbf{u}_G= -\nabla p
\end{equation}
where $\mathbf{u}_G=U_G \hat{i}$.
The appropriate boundary conditions for the PBL are that the velocity vanishes at the lower boundary and approaches the geostrophic velocity, $U_G$, as the height tends to infinity.
\begin{equation}
\mathbf{u}(y=0)=0, \mathbf{u}(y=\infty)=U_G\hat{i}
\end{equation}\\
In the absence of background turbulence, a laminar equilibrium in the PBL is maintained by a three way momentum balance between diffusion, the Coriolis force, and the pressure gradient force:\\
\begin{equation}
2 \Omega \hat{y} \times \mathbf{u}_e =-\nabla p+ \nu \Delta \mathbf{u}_e
\end{equation}\\
The resulting profile, commonly referred to as the laminar Ekman Layer, is written as:\\
\begin{equation}U_e(y)=U_G(-e^{-y/D}cos(\frac{y}{D})+1)\end{equation}
\begin{equation}V_e(y)=0\end{equation}
\begin{equation}W_e(y)=U_G(-e^{-y/D}sin(\frac{y}{D}))\end{equation}\\
where $D=\sqrt{\frac{\nu}{\Omega}}$ defines the characteristic length scale of the PBL.\\ 

The dimensionless numbers are:\\
\begin{equation}
Re=\frac{U_GD}{\nu}=\frac{U_G}{\sqrt{\nu \Omega}},
Ro=\frac{U_G}{2\Omega D}=\frac{1}{2}\frac{U_G}{\sqrt{\nu \Omega}}=\frac{1}{2}Re\end{equation}\\

The dimensionless governing equation is:\\
\begin{equation}\frac{\partial \mathbf{u}}{ \partial t}+(\mathbf{u}\cdot \nabla)\mathbf{u}=- \nabla p-\frac{2}{Re}\hat{y} \times \mathbf{u} + \frac{1}{Re}\Delta \mathbf{u} \label{dimlessfulleq}\end{equation}\\

The resulting equilibrium laminar Ekman layer profile is:\\ 
\begin{equation}U_e(y)=-e^{-y}cos(y)+1\end{equation}
\begin{equation}V_e(y)=0\end{equation}
\begin{equation}W_e(y)=-e^{-y}sin(y)\end{equation}\\

We impose periodic boundary conditions in both the streamwise ($x$) and spanwise ($z$) directions. In the wall-normal direction ($y$), we apply a stress-free boundary condition at the top and a no-slip boundary condition at the bottom.

In anticipation of RSS formation, the coordinate system is rotated counterclockwise by $\theta = \frac{15}{180}\pi$ (15 degrees), consistent with the $O(10^{\circ} - 30^{\circ})$ deviation of the roll axis from the geostrophic direction observed in both measurements \citep{LeMone1973,Foster2013,Gerling1986,Weckwerth1999,Belusic2015,Tang2021} and numerical simulations \citep{LappenRandall2005,Dubos2008,Esau2012}. Unless explicitly stated otherwise, analyses henceforth are performed in this rotated coordinate system.\\

The Ekman layer solution in the rotated coordinate system is:\\ 
\begin{equation}U_e(y)=cos(\theta)-e^{-y}cos(y+\theta)
\label{ekmanspiralU}
\end{equation}
\begin{equation}V_e(y)=0\label{ekmanspiralV}\end{equation}
\begin{equation}W_e(y)=sin(\theta)-e^{-y}sin(y+\theta)\label{ekmanspiralW}\end{equation}\\

\section{Formulating the SSD Equations for the PBL}
To derive the SSD equations, the variables are decomposed into their mean and fluctuating parts. Accordingly, the velocity field $\mathbf{u}$ is written as the sum of its mean component and its fluctuation:\\
\begin{equation}
\mathbf{u}(x,y,z,t)=\mathbf{U}(y,z,t)+\mathbf{u'}(x,y,z,t)
\end{equation}\\

Using an overbar to denote the Reynolds averaging operator and capital letters to represent Reynolds-averaged variables, we define our Reynolds average in a coordinate system where the x-direction is rotated 15° counterclockwise from the geostrophic wind direction.\\
\begin{equation}
\bar{\mathbf{u}}=[\mathbf{u}]_x=\mathbf{U}
\end{equation}\\

By choosing a direction for the Reynolds average that is rotated away from the geostrophic, we anticipate that the RSS instability will break the model’s symmetry in the corresponding spanwise direction, and that this instability and its finite-amplitude equilibria can, through this choice, be isolated within the first cumulant. The equations for the mean velocity $\mathbf{U}$ are then obtained by taking the streamwise average of equation \eqref{dimlessfulleq}.\\

\begin{equation}\frac{\partial U}{\partial t}=U_y\Psi_z -U_z\Psi_y-\partial_y\overline{u'v'}- \partial_z\overline{u'w'}+\Delta_1 \frac{U}{Re}-\frac{2 }{Re}\Psi_y +\frac{2}{Re}sin(\theta)
\label{meanUeq}
\end{equation}
\begin{equation}\frac{\partial (\Delta_1 \Psi)}{\partial t}=(\partial_{yy} - \partial_{zz})(\Psi_y \Psi_z - \overline{v'w'})-\partial_{yz}(\Psi_y^2-\Psi_z^2+\overline{w'^2}-\overline{v'^2})+\Delta_1 \Delta_1 \frac{\Psi}{Re}+\frac{2}{Re}U_y 
\label{meanPsieq}
\end{equation}\\
in which $\Delta_1:=(\partial_{yy}+\partial_{zz})$ and where we have taken advantage of nondivergence in the spanwise/cross-stream plane to write We represent the spanwise and cross-stream velocities using the streamfunction as
\(W = \frac{\partial \Psi}{\partial y}\) and \(V = -\frac{\partial \Psi}{\partial z}\). By expressing the streamwise mean velocity component in terms of \(U\) and \(\Psi\), the Reynolds stress structures are confined to the first cumulant.\\

 The fluctuation velocity dynamics is formulated in terms of the wall-normal velocity and wall-normal vorticity, a representation in which the nondivergence of the velocity field is inherently satisfied \citep{Schmid-Henningson-2001}. The equations for the fluctuating velocity components, $\mathbf{u}'$, are obtained by subtracting the mean-flow equations \eqref{meanUeq} and \eqref{meanPsieq} from the corresponding full equations \eqref{dimlessfulleq} component by component.\\ 

Fluctuation wall normal velocity and vorticity are decomposed into streamwise wavenumber components as 
\begin{equation}v'(x,y,z,t)=\sum_{k}^{}v'_k(y,z,t)e^{ikx}\end{equation}
\begin{equation}\eta'(x,y,z,t)=\sum_{k}^{}\eta'_k(y,z,t) e^{ikx}\end{equation}

We can now incorporate the fluctuation velocity equations in the wall normal velocity-vorticity formulation into our perturbation equations to obtain these equations in the compact form:\\

\begin{equation}\frac{\partial \mathbf{\phi}_k}{\partial t}=A \mathbf{\phi}_k+\mathbf{\xi}_k\end{equation} \\
in which the fluctuation state vector is: \\
 \begin{equation}
 \mathbf{\phi}_k=
 \begin{bmatrix} v'_k \\ \eta'_k \end{bmatrix}
 \end{equation}\\
and we have parameterized the fluctuation-fluctuation nonlinear term by a stochastic noise process $\xi(t)$.
We note that in applying this model to physical problems, this stochastic noise process incorporates also any external sources of turbulent fluctuations.\\

The matrix of the dynamics is:
 \\
 \begin{equation}A=\begin{bmatrix}
       A_{11} & A_{12} \\ A_{21} & A_{22}
        \end{bmatrix}\end{equation}
        \\
        \begin{equation}A_{11}=L_{OS}(U)+\Delta^{-1} (LV_{11}(V)+LW_{11}(W)) \label{A11eq} \end{equation}
        \begin{equation}A_{12}=L_{C1}(U)+\Delta^{-1}(LV_{12}(V)+LW_{12}(W))-\Delta^{-1}(\frac{2}{Re}\partial_y)
        \label{A12eq}\end{equation}
        
        \begin{equation}A_{21}=L_{C2}(U)+LV_{21}(V)+LW_{21}(W)+\frac{2}{Re}\partial_y
        \label{A21eq}
        \end{equation}
        \begin{equation}A_{22}=L_{SQ}(U)+LV_{22}(V)+LW_{22}(W)
        \label{A22eq}
        \end{equation}
       
where $\Delta_2=(\partial_{xx}+\partial_{zz})$. \\

$L_{OS}(U)$ denotes the Orr–Sommerfeld operator, where $U$ is the streamwise mean velocity. For a detailed description of the Orr–Sommerfeld–Squire formulation of the dynamics, see~\citet{Farrell-Ioannou-2012}. The individual components of Eqs.~\eqref{A11eq}, \eqref{A12eq}, \eqref{A21eq}, and \eqref{A22eq} are described in Appendix~\ref{appendix:Adetail}.\\
Following the SSD formulation presented in~\citet{Farrell2019}, one can derive a deterministic Lyapunov equation for the ensemble-mean fluctuation covariance $C$, using only the operator $A$ and a white-in-time stochastic forcing with spatial covariance $Q := \langle \xi \xi^\dagger \rangle$.\\ 

\begin{equation}\frac{\partial 
\mathbf{C}_k}{\partial t}=A \mathbf{C}_k + \mathbf{C}_k A^{\dagger}+ \epsilon \mathbf{Q} \end{equation}
\begin{equation}\mathbf{C}_{k}=\mathbf{\phi}_k \mathbf{\phi}_k^{\dagger} \end{equation}\\
where $\dagger$ denotes the Hermitian transpose. We emphasize that it is the existence of this time-dependent Lyapunov equation that allows us to obtain a deterministic SSD.\\

The Reynolds stress terms appearing in the mean equations \eqref{meanUeq} \eqref{meanPsieq} can be obtained by applying a linear operator, $\mathbf{L}_{RS}$, to the covariances, $\mathbf{C}_{k}$, of the fluctuations cf.~\citep{Farrell-Ioannou-2012}.  Taking account of this, the equations for the mean state can be expressed in compact form as:\\

\begin{equation}  \frac{\partial\bv{\Gamma}}{\partial t}=\mathbf{G}(\bv{\Gamma})+\sum_{k}^{} \mathbf{L}_{RS}\mathbf{C}_{k} \end{equation}\\
where $\bv\Gamma=[U, \Psi]^T$ and $\bv{G}$ is defined as:\\
\begin{equation}\mathbf{G}(\bv{\Gamma})=\begin{bmatrix}
(U_y)\Psi_z - U_z \Psi_y+\Delta_1 \frac{U}{Re}-\frac{2}{Re}\Psi_y+\frac{2}{Re}sin(\theta)\\
\Delta_1^{-1}[(\partial_{yy} - \partial_{zz})(\Psi_y \Psi_z)-\partial_{yz}(\Psi_y^2-\Psi_z^2)+\Delta_1 \Delta_1 \frac{\Psi}{Re}+\frac{2}{Re}U_y]
\end{bmatrix}\end{equation}\\
and the forcing by the fluctuation stresses at each $k$ is:\\

\begin{equation}\mathbf{L}_{RS}\mathbf{C}_{k}=\begin{bmatrix}
- \partial_y \overline{u'v'}|_{k}- \partial_z \overline{u'w'}|_{k}\\
\Delta_1^{-1}[(\partial_{yy} - \partial_{zz})( - \overline{v'w'}|_{k})-\partial_{yz}(\overline{w'^2}|_k-\overline{v'^2}|_k)]
\end{bmatrix}\end{equation}\\
in which the fluctuation stress operator $\mathbf{L}_{RS}$ has been composed using these linear operators:\\

\begin{equation}
\begin{aligned}
\mathbf{L}_{u'}^{k}&=[-ik \Delta_2^{-1} \partial_y , \Delta_2^{-1} \partial_z ]&&\\
\mathbf{L}_{v'}^{k}&=[I, 0]&&\\
\mathbf{L}_{w'}^{k}&=[-\Delta_2^{-1} \partial_{yz}, -ik \Delta_2^{-1}].&&\\
\end{aligned}
\end{equation}\\

For instance, when a grid based method is used in both  y and  z:\\

\begin{equation}\overline{u'v'}|_k=diag(\mathbf{L}_{u'}^{k}\mathbf{C}_k \mathbf{L}_{v'}^{k\dagger})\end{equation}\\
and the other stress terms are written in a similar manner.  Note that stress terms $\sum_{k}\mathbf{L}_{RS}\mathbf{C}_k$ are linear in $\mathbf{C}_k$. \\

For our RSS stability analysis, we employ a second-order closed SSD formulation known as S3T. This SSD framework comprises the first and second cumulants, supplemented by a stochastic closure. The S3T equations can be expressed compactly as:\\

\begin{equation}
\frac{\partial \bv{\Gamma}}{\partial t}=\mathbf{G}(\bv\Gamma)+\sum_{k}^{} \mathbf{L}_{RS}\mathbf{C}_{k}
\label{SSDmeaneq}
\end{equation}
\begin{equation}
\frac{\partial \mathbf{C}_k}{\partial t}=\mathbf{A} \mathbf{C}_k + \mathbf{C}_k \mathbf{A}^{\dagger}+ \epsilon \mathbf{Q} 
\label{SSDcoveq}
\end{equation}

Although many choices for $Q$ are possible, we aim to minimally affect the dynamics and therefore choose it to be white in kinetic energy, i.e., to excite each degree of freedom equally in kinetic energy while reflecting the physically expected decay of background turbulence with distance from the lower boundary.  This is accomplished by choosing $\mathbf{Q}$ as follows \citep{Farrell-Ioannou-2012}:
\begin{equation}\mathbf{Q}=S_{D}(y)\cdot \mathbf{M}^{-1}\end{equation}
\begin{equation}
S_D(y)=\frac{1}{2} (\left(\tanh\left(y+2\pi\right)-\tanh\left(y-2\pi\right)\right)
\end{equation}
\begin{equation}\mathbf{M}=(\mathbf{L}_{u'}^{k\dagger}\mathbf{L}_{u'}^{k}+\mathbf{L}_{v'}^{k\dagger}\mathbf{L}_{v'}^{k}+\mathbf{L}_{w'}^{k\dagger}\mathbf{L}_{w'}^{k})/(2*Ny*Nz)\end{equation}\\

This completes the formulation of the SSD. We now turn our attention to analyzing the stability of this system.

\section{S3T Stability Formulation}
In order to study stability, we first need to establish equilibrium states for which the time derivative of equations \eqref{SSDmeaneq} and \eqref{SSDcoveq} vanish:

\begin{equation}
\bv{\Gamma}_e=\begin{bmatrix} U_e\\\Psi_e\end{bmatrix}\end{equation}
\begin{equation}
\mathbf{C}_e=\sum_k\mathbf{C}_{ke}
\end{equation}\\
 
The stability properties of the SSD equilibria, $\bv{\Gamma}_e$ and $\sum_k \bv{C}_{ke}$, are obtained by linearizing \eqref{SSDmeaneq} and \eqref{SSDcoveq} about these equilibrium states. Setting the background turbulence parameter to $\epsilon = 0$ recovers the laminar Ekman layer spiral solutions \eqref{ekmanspiralU}, \eqref{ekmanspiralV}, and \eqref{ekmanspiralW}. As $\epsilon$ increases, the equilibrium velocity $U_e(y)$ and streamfunction $\Psi_e(y)$ deviate from the laminar Ekman spiral. Thus, the SSD equilibrium state $(\bv{\Gamma}_e, \sum_k \bv{C}_{ke})(\epsilon)$ depends on the background turbulence intensity parameter $\epsilon$. 
Linear perturbation equations linearized around the SSD equilibrium state  ($\bv{\Gamma}_e$, $\sum_k\bv{C}_{ke}$) are:

\begin{equation}\frac{\partial (\delta \bv{\Gamma})}{\partial t}=\sum_{i}^{} \frac{\partial \bv{G}}{\partial \Gamma_i}|_{\bv{\Gamma}_{e}} \delta \Gamma_i+ \sum_{k}^{}\bv{L}_{RS}\delta \bv{C}_k 
\label{SSDinstabilitymeaneq}
\end{equation}
\begin{equation}\frac{\partial (\delta \bv{C}_k)}{\partial t}=\bv{A}_{keq}\delta \bv{C}_k+ \delta \bv{C}_k \bv{A}_{keq}^{\dagger}+ \delta \bv{A}_{k} \bv{C}_{keq}+\bv{C}_{keq}\delta \bv{A}_{k}^{\dagger}
\label{SSDinstabilitycoveq}
\end{equation}
where 
\begin{equation}\delta \bv{A}_k=\bv{A}_k(\bv{\Gamma}_e+ \delta \bv{\Gamma})-\bv{A}_k(\bv{\Gamma}_e).\end{equation}\\

Equations \eqref{SSDinstabilitymeaneq} and \eqref{SSDinstabilitycoveq} comprise the formulation of the linear perturbation S3T dynamics. 
We set the Reynolds number of our PBL model to $Re = 300$, leaving the background turbulence excitation intensity, $\epsilon$, as the only adjustable parameter. The matrix $Q$ is scaled so that $\epsilon = 1$ corresponds to a volume-averaged RMS perturbation velocity equal to $1\%$ of the maximum geostrophic velocity magnitude. Specifically, $Q$ is chosen such that $\sqrt{2\langle E_k \rangle} = 0.01$, where $\langle E_k \rangle = \operatorname{trace}(M_k C_k)$ is the ensemble-averaged kinetic energy density of the perturbation field, following \citet{Farrell-Ioannou-2012}.

Analysis of this SSD stability problem uncovers an RSS instability that is intrinsically different from standard hydrodynamic instabilities. Although the S3T framework does include laminar modal instabilities through its first cumulant (the mean flow), its central contribution is to reveal a manifold of instabilities produced by the interaction between the mean flow and the fluctuations. These are governed by equations \eqref{SSDinstabilitymeaneq} and \eqref{SSDinstabilitycoveq}, which define the first and second cumulants of our S3T SSD. Mechanistically, this manifold of RSS modes becomes unstable because, in the presence of a streamwise streak, the background turbulence is systematically strained so that the resulting Reynolds stresses exert a torque that drives a roll circulation aligned with the original streak \citep{Farrell 2022, Nikolaidis 2024}.
The S3T SSD formulation of the PBL stability problem incorporates both classical inflectional instability and the instability mechanism driven by Reynolds stress (RS) torque feedback. When the mean equation \eqref{SSDmeaneq} is considered in isolation from the covariance equation, it recovers the familiar inflectional instability dynamics of the Ekman layer \citep{Lilly1966, FallerKaylor1966}. The RS instability mechanism has previously been revealed through the S3T framework, which identified RSS instabilities arising solely from the interaction between the mean flow and fluctuations in unstratified wall-bounded shear flows~\citep{Farrell-Ioannou-2012, Farrell-Ioannou-2016-bifur}. A full understanding of RSS dynamics in the turbulent PBL, however, requires incorporating both the classical inflectional instability captured by the mean flow equation and the RS torque mechanism that results from coupling the mean flow equation with the fluctuation covariance equation.
\section{RSS Instability in the S3T SSD Implementation of Turbulent PBL Dynamics}

The Ekman-layer spiral supports inflectional instabilities that have been associated with the formation and equilibration of RSS in the Ekman boundary layer. To assess the influence of these modes, we suppress them here so as to isolate the impact of the RS-torque instability mechanism on the dynamics of PBL RSS. Similar strategies for suppressing laminar modal instabilities have been used in several earlier studies of turbulent roll–streak dynamics\citep{Duran2019,Duran2021,Kimeadypart1,Kimeadypart2}.

S3T dynamics with laminar modal instabilities suppressed is expressed as:\\
\begin{equation}
\frac{\partial (\delta \bv{\Gamma})}{\partial t}=\sum_{i}[\frac{\partial \mathbf{G}}{\partial \Gamma_i}(\lambda<0)]_{\bv{\Gamma}_e} \delta \Gamma_i+ \sum_{k}\mathbf{L}_{RS}\delta \mathbf{C}_k
\label{nolammodemeanS3Teq}
\end{equation}
\begin{equation}
\frac{\partial (\delta \mathbf{C}_k)}{\partial t}=\mathbf{A}_{ke}(\lambda <0) \delta \mathbf{C}_k+ \delta \mathbf{C}_k \mathbf{A}_{ke}(\lambda <0)^{\dagger}+\delta \mathbf{A}_k \mathbf{C}_{ke}+ \mathbf{C}_{ke}\delta \mathbf{A}_k^{\dagger}
\label{nolammodecovS3Teq}
\end{equation}\\
where terms in which modal instabilities have been suppressed are indicated by $[\frac{\partial \mathbf{G}}{\partial \Gamma_i}(\lambda<0)]_{\bv{\Gamma}_e}$, $\mathbf{A}(\lambda <0)_{ke}$, and $\mathbf{A}(\lambda <0)_{ke}^{\dagger}$.

\begin{figure}
\centering{
\begin{subfigure}{0.8\textwidth} 
\centering{\caption{}\includegraphics[width=0.75\linewidth]{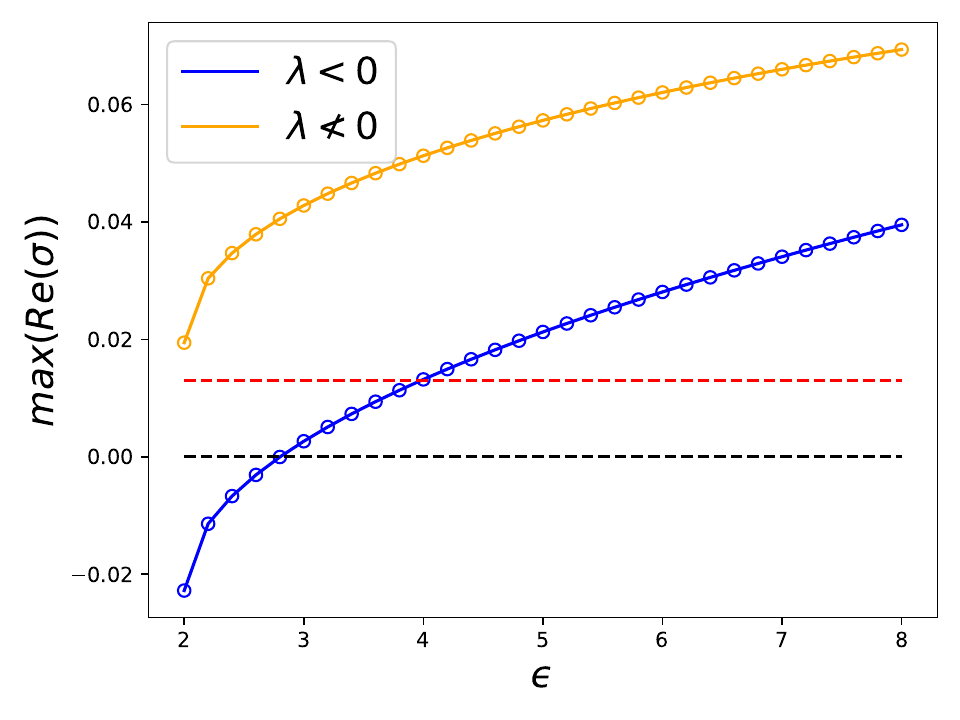}
\label{fig:1a}} 
\end{subfigure}
\begin{subfigure}{0.8\textwidth}\centering{ \caption{}
\includegraphics[width=0.75\linewidth]{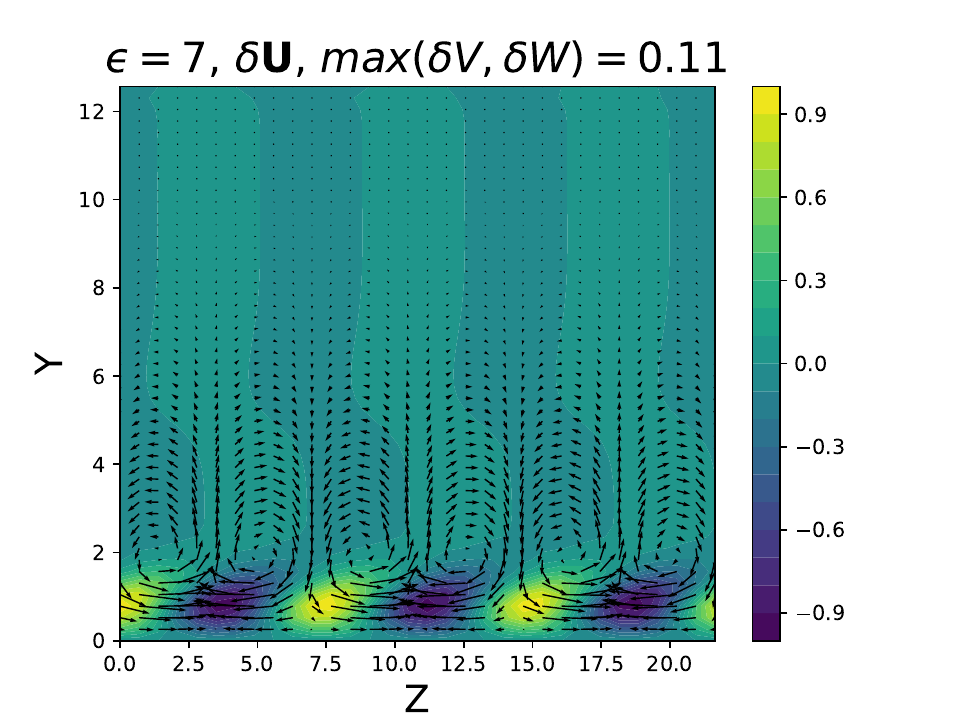} \label{fig:1b}} 
\end{subfigure}

\caption{(a) Stability diagram depicting the maximum growth rate, $\max(\text{Re}(\sigma))$, as a function of the background turbulence intensity parameter $\epsilon$ for both suppressed ($\lambda < 0$) and unsuppressed ($\lambda \ge 0$) inflectional-mode instabilities. The black dashed line marks the stability boundary, while the red dashed line represents the growth rate of the pure inflectional mode ($\epsilon = 0$). (b) Streamwise velocity contours and spanwise/cross-stream velocity vectors of the most unstable eigenmode in the unsuppressed case. The eigenfunctions are normalized such that the maximum value of $\delta U$ equals unity. Results are shown for $\epsilon = 7$; $Re = 300$.}
 \label{fig:instability}}
\end{figure}
Shown in blue line of panel $(a)$ of figure \ref{fig:instability} is the stability diagram as a function of turbulence excitation parameters $\epsilon$ for the turbulent Ekman layer obtained using the linear perturbation form of the S3T SSD with suppression of inflectional instability as indicated in \eqref{nolammodemeanS3Teq} \eqref{nolammodecovS3Teq}. 
 Black dashed line indicates the line of neutral stability. 
The red line indicates the growth rate of the inflectional mode.  This stability diagram reveals that the RS S3T RSS mode growth rate becomes equal to that of the inflectional mode for the modest intensity of background turbulence supported by $\epsilon\approx 4.0$. And, growth rate continues to increase beyond the growth rate of the inflectional mode when $\epsilon$ increases further. To understand the potential interaction between these two mechanisms, we now unsuppress inflectional instability and go back to S3T SSD implementation of turbulent PBL, equations  \eqref{SSDinstabilitymeaneq} and \eqref{SSDinstabilitycoveq}, containing both inflectional instability mechanism and RS torque mechanism. Orange line of the panel $(a)$ shows the stability diagram as a function of turbulence excitation parameters $\epsilon$ for the turbulent Ekman layer obtained using the linear perturbation form of the S3T SSD without suppression of inflectional instability. 
These mechanisms initially act in synergy to destabilize the laminar Ekman profile as indicated by their combined growth rate being greater than the maximum of their individual growth rates.

In this section, we used linear perturbation analysis of the S3T SSD to study the RSS instability in the context of PBL. In the next section, we turn our focus to the equilibration of these instabilities in the non-linear S3T SSD.

\section{Statistical Equilibrium States in the S3T SSD Implementation of Turbulent PBL Dynamics}
\begin{figure}
\centering{
\includegraphics[width=\linewidth]{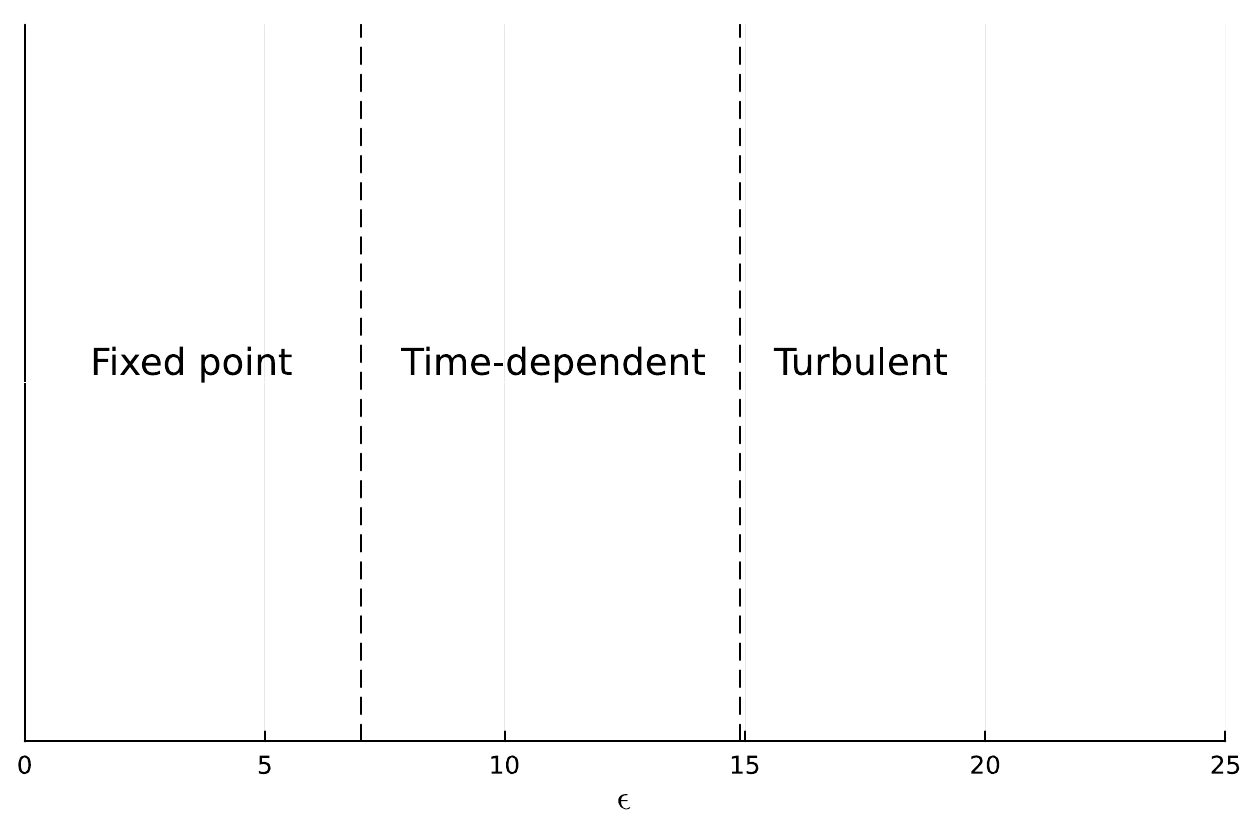}}
\caption{Regime diagram showing the equilibrium RSS state as a function of the background turbulence intensity parameter, $\epsilon$, which controls the strength of the RS torque mechanism. A weak inflectional instability is present, allowing a fixed point to exist for $0 \leq \epsilon \leq 6.9$. For $\epsilon > 6.9$, the system transitions to time-dependent behavior. At higher values, $\epsilon > 14.9$, a turbulent RSS regime emerges; $Re = 300$.  }
\label{nobetaequilibrium}
\end{figure}

Although identifying an instability mechanism with eigenstructures in the form of a RSS suggests that the observed RSS may originate from this mechanism, firmly linking an unstable perturbation RSS to the finite-amplitude structure of the observed RSS requires extending the analysis to the nonlinear equilibration governed by the full S3T SSD equations, \eqref{SSDmeaneq} and \eqref{SSDcoveq}.
Depending on the parameter regime, instabilities can equilibrate to fixed-point, time-dependent, or turbulent states. Moreover, the turbulent state can be self-sustaining in the sense that the background turbulence excitation parameter, $\epsilon$, may be set to zero while the turbulence continues to be maintained. To summarize our results and set the stage for their presentation, an equilibrium structure diagram as a function of $\epsilon$ is shown in figure \ref{nobetaequilibrium}.\\

\begin{figure}
\centering{
\includegraphics[width=0.75\linewidth]{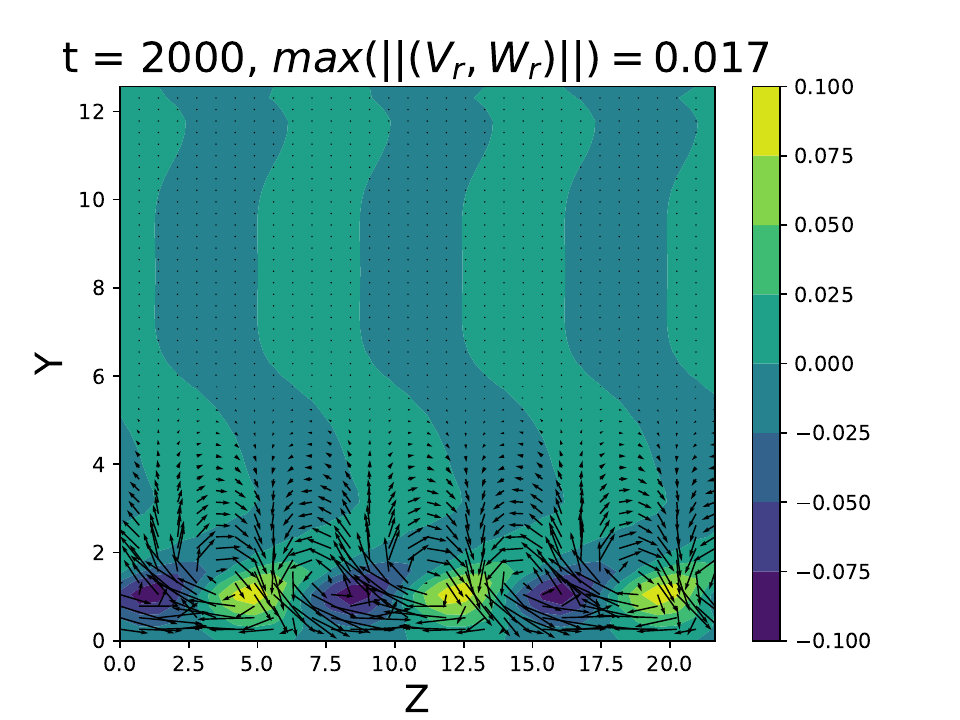}}
\caption{Fixed-point equilibrium of the RSS system, exhibiting finite-amplitude contributions from both the streak, $U_s$, and the roll velocity, $(V_r, W_r)$. The streak is defined as $U_s = U - [U]_z$, and the roll components as $V_r = V - [V]_z$ and $W_r = W - [W]_z$. The streak is shown in color, while the roll velocity is represented by vectors. Parameters: $\epsilon = 2$; $Re = 300$. }
\label{fixedpointsnapshot}
\end{figure}

We now examine several dynamically significant regions of this diagram.
Finite-amplitude RSS state regimes are traditionally distinguished by examining time series of the fluctuation energy:\\
\begin{equation}
TKE(t)=\frac{1}{2}[u'^2+v'^2+w'^2]_{x,y,z}
\end{equation}\\
and by examining snapshots of structure.

\begin{figure}
\centering{
\includegraphics[width=\linewidth,height=0.5\linewidth]{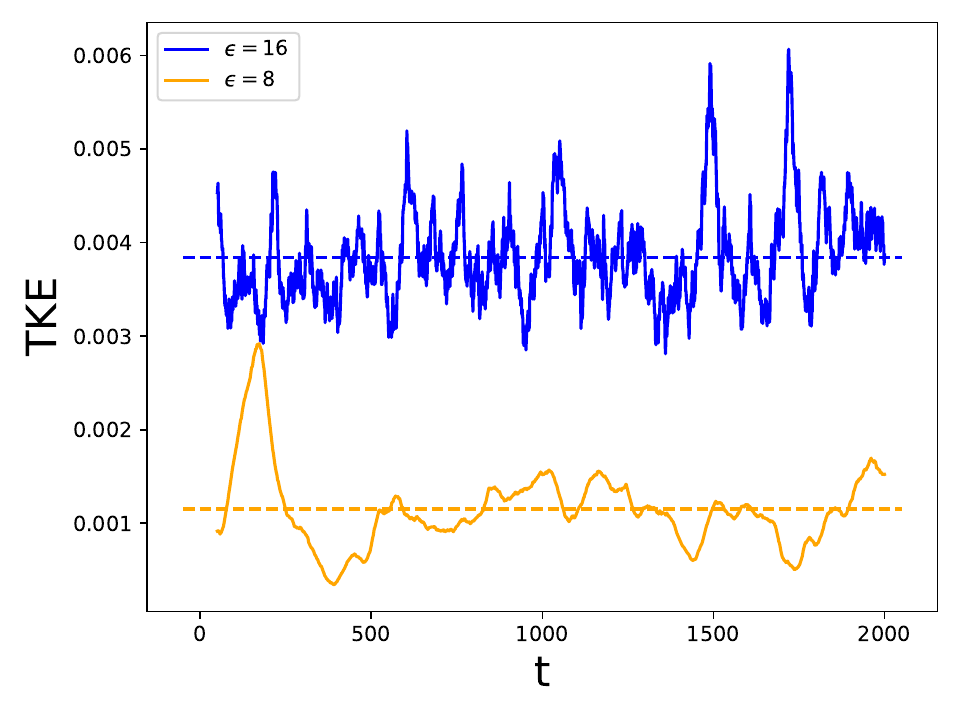}
}
\caption{Time series of fluctuating TKE for a time-dependent (orange line) and a turbulent equilibrium (blue line).The time-dependent equilibrium is at $\epsilon = 8$, while the turbulent equilibrium is at $\epsilon = 16$; $Re = 300$.}
\label{tketimedepvsturb}
\end{figure}

\begin{figure}

\subfloat[]{%
            \includegraphics[width=.48\linewidth]{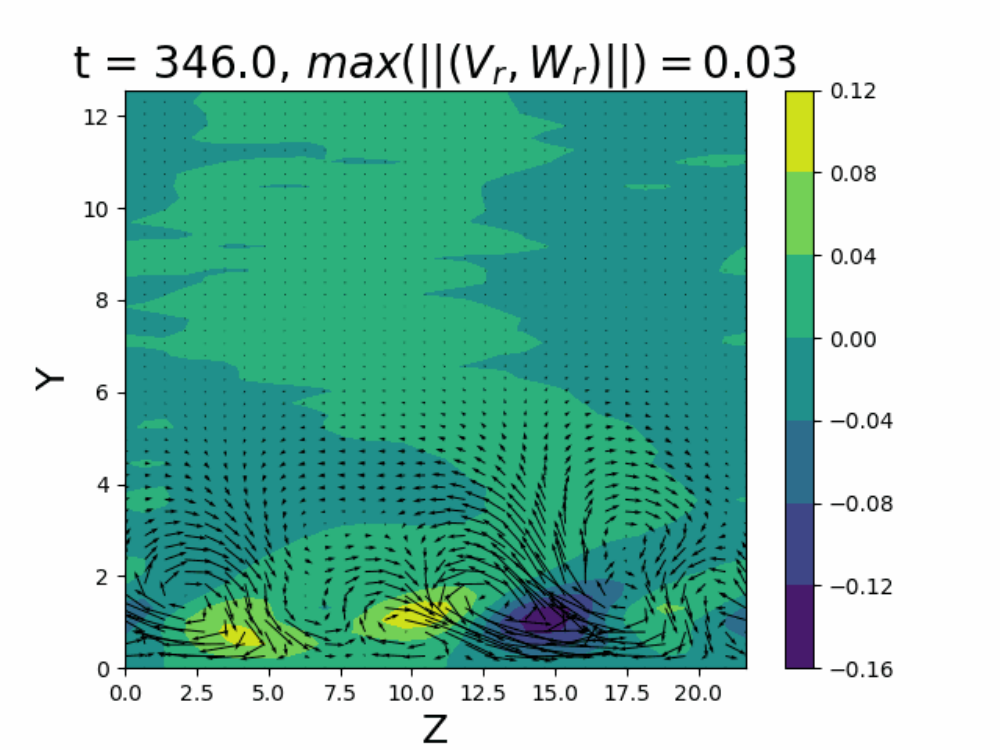}
            \label{subfig:a}%
        }\hfill
        \subfloat[]{%
            \includegraphics[width=.48\linewidth]{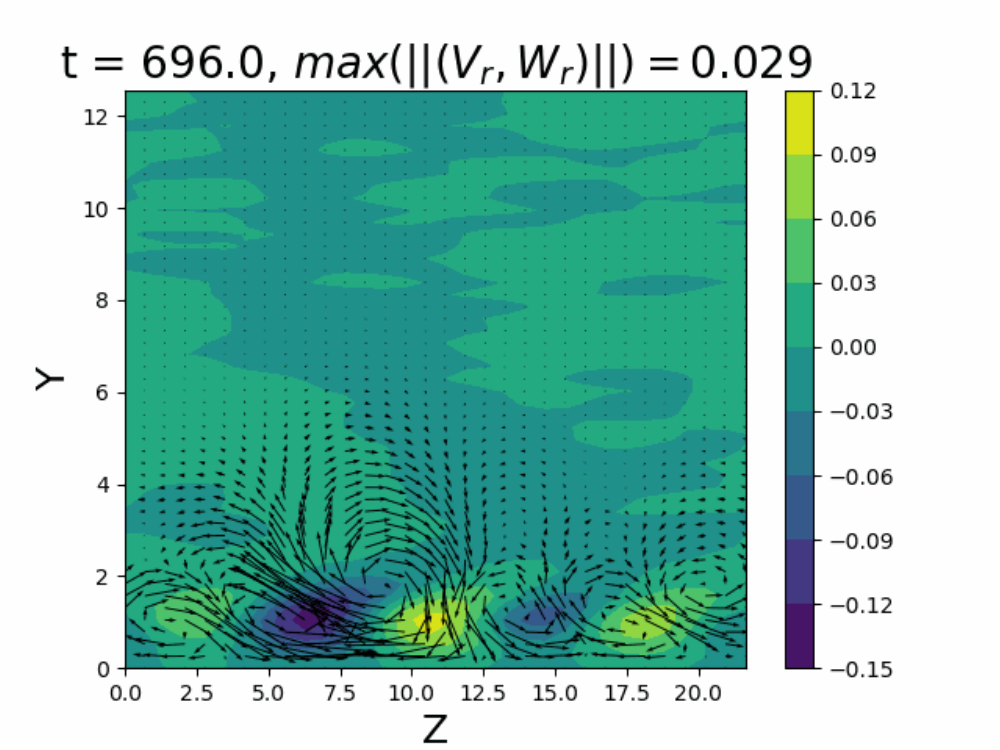}
            \label{subfig:b}%
        }\\
        \subfloat[]{%
            \includegraphics[width=.48\linewidth]{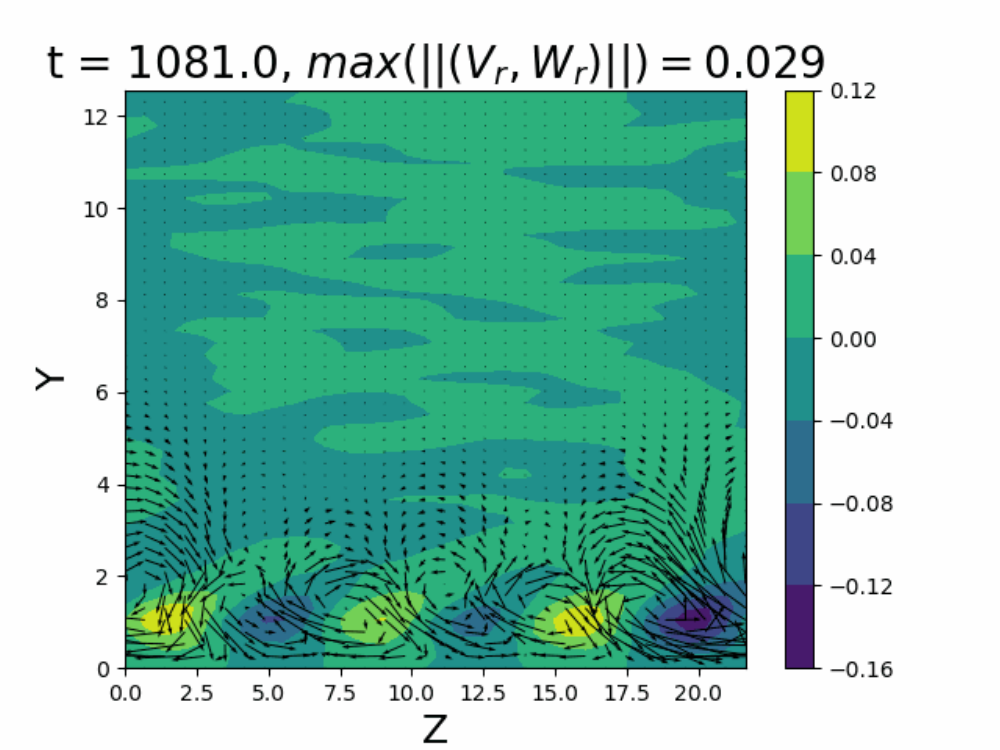}
            \label{subfig:c}%
        }\hfill
        \subfloat[]{%
            \includegraphics[width=.48\linewidth]{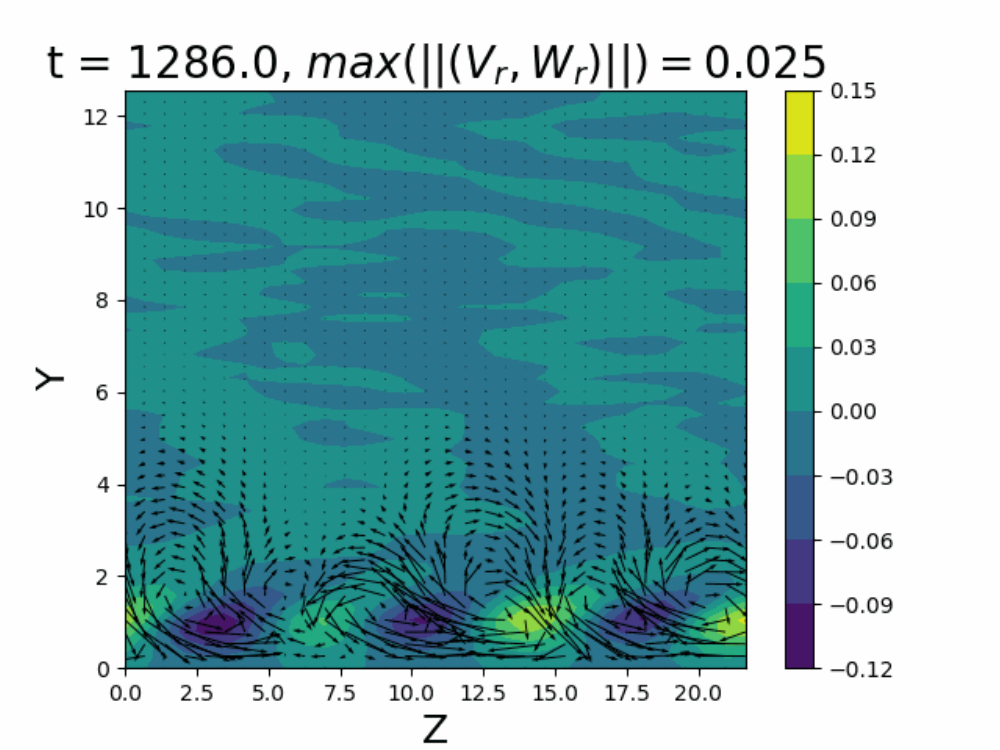}
            \label{subfig:d}%
        }
\caption{Snapshots of a time-dependent equilibrium. The streak is shown in color, and the roll velocity
is represented by vectors. Parameters: $\epsilon = 8$, $Re = 300$. }
\label{timedepsnapshots}
\end{figure}

\begin{figure}

\subfloat[]{%
            \includegraphics[width=.48\linewidth]{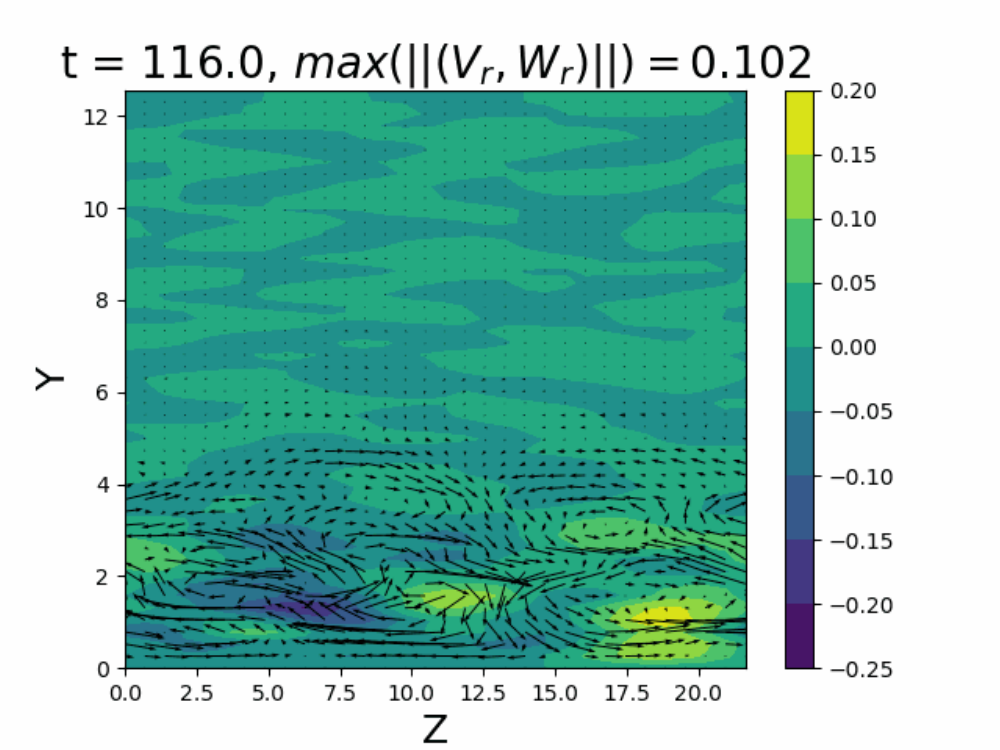}
            \label{subfig:a}%
        }\hfill
        \subfloat[]{%
            \includegraphics[width=.48\linewidth]{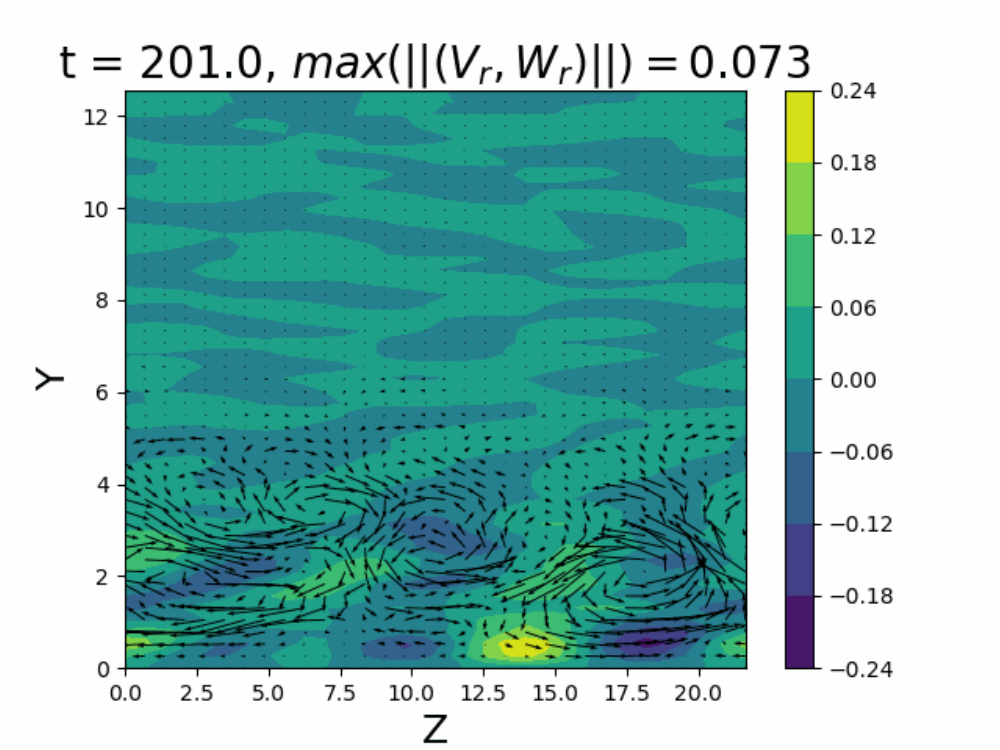}
            \label{subfig:b}%
        }\\
        \subfloat[]{%
            \includegraphics[width=.48\linewidth]{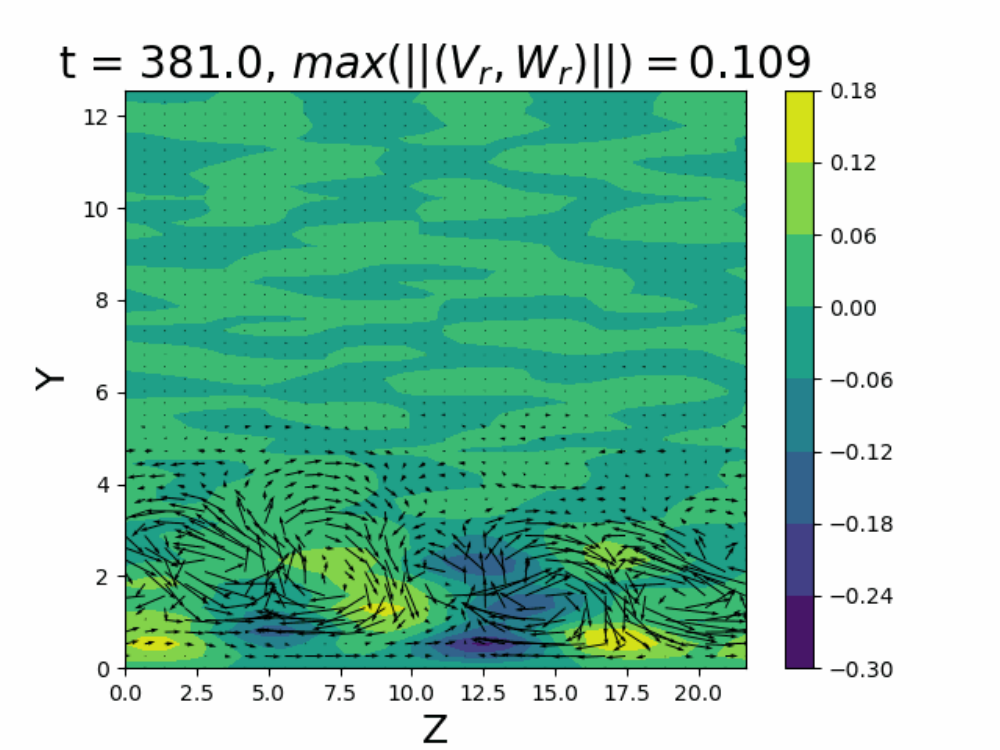}
            \label{subfig:c}%
        }\hfill
        \subfloat[]{%
            \includegraphics[width=.48\linewidth]{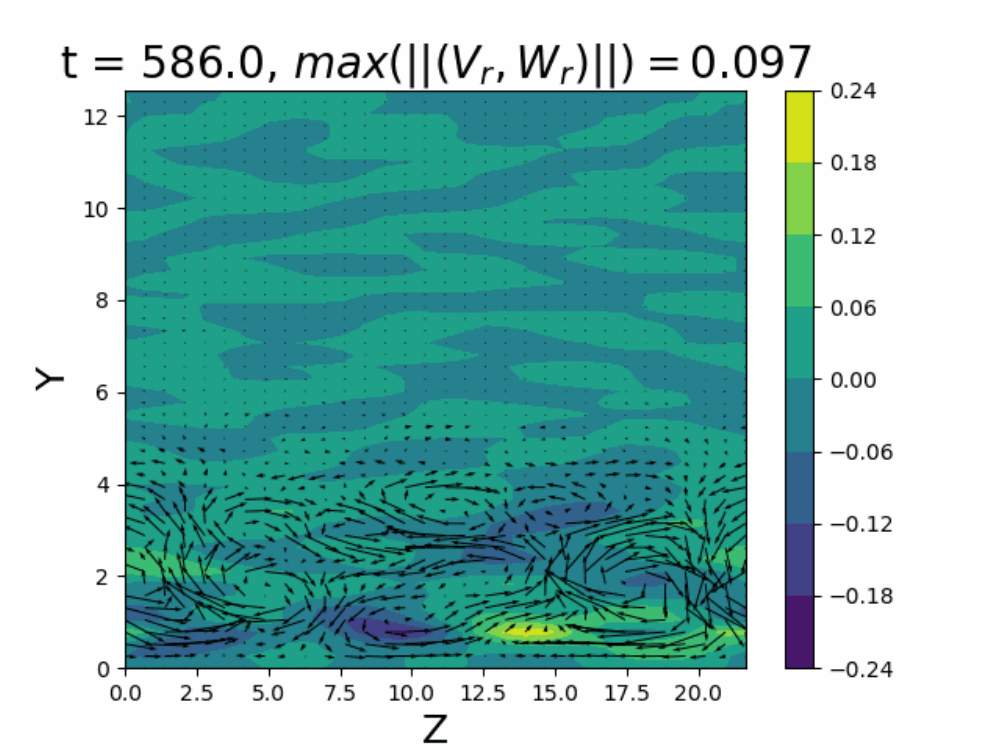}
            \label{subfig:d}%
        }
\caption{Snapshots of a turbulent equilibrium. The streak is shown in color, and the roll velocity
is represented by vectors. Parameters: $\epsilon = 16$, $Re = 300$. }
\label{turbulentsnapshots}
\end{figure}

\begin{figure}
\centering{\begin{subfigure}{0.8\textwidth}\caption{}
\includegraphics[width=\linewidth,height=0.5\linewidth]{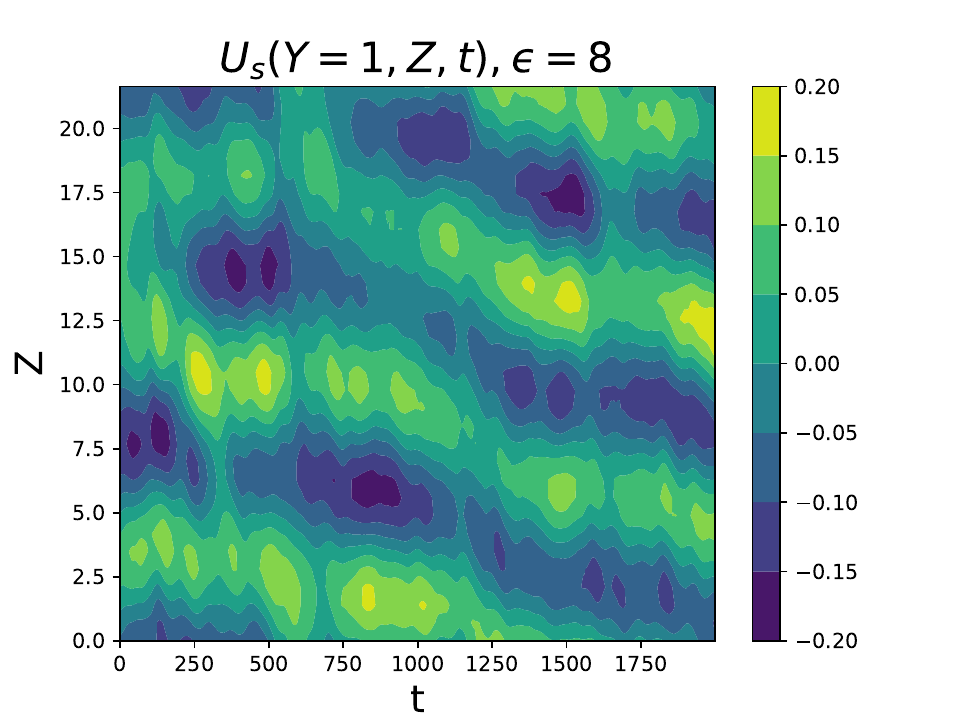}
\end{subfigure}
\begin{subfigure}
{0.8\textwidth}\caption{}
\includegraphics[width=\linewidth,height=0.5\linewidth]{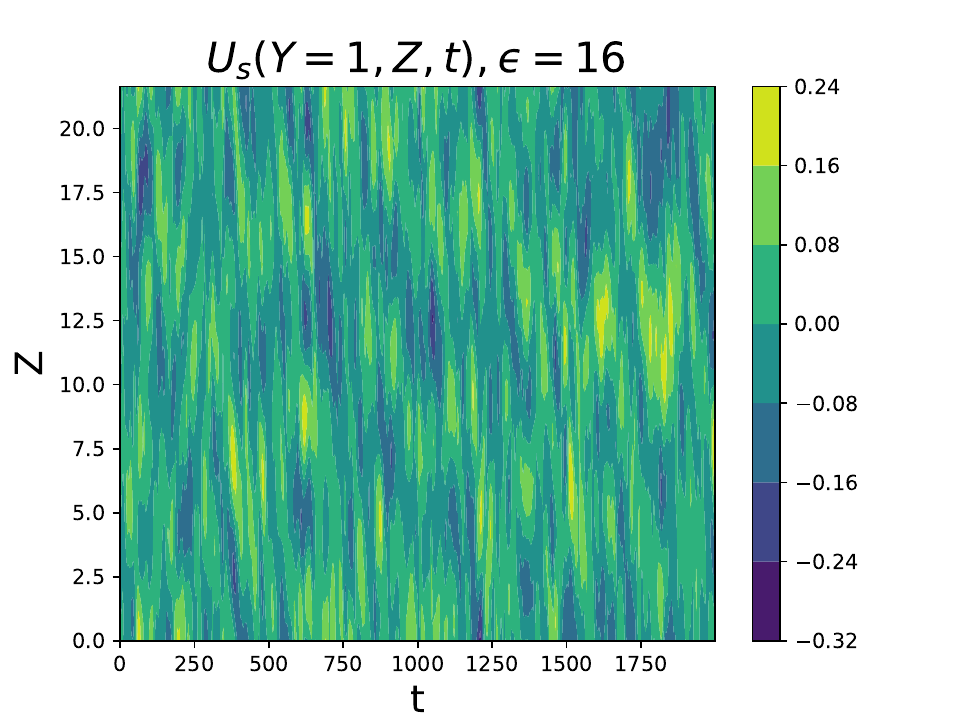}
\end{subfigure}
}
\caption{Panel $(b)$: Hovmöller diagram of $U_s(y=1,z,t)$ for a time-dependent equilibrium. Panel $(c)$: Hovmöller diagram of $U_s(y=1,z,t)$ for a turbulent equilibrium. The time-dependent equilibrium is at $\epsilon = 8$, while the turbulent equilibrium is at $\epsilon = 16$; $Re = 300$.}
\label{Hovmollertimedepvsturb}
\end{figure}

Fixed-point S3T SSD equilibria are characterized by time-independent TKE and a stationary flow structure, whereas both time-dependent and turbulent equilibria exhibit TKE and structures that vary in time. A snapshot of a fixed-point equilibrium is shown in figure \ref{fixedpointsnapshot}. Figure \ref{tketimedepvsturb} presents TKE time series for a time-dependent equilibrium ($\epsilon = 8$) and a turbulent equilibrium ($\epsilon = 16$). Snapshots of the time-dependent equilibrium ($\epsilon = 8$) appear in figure \ref{timedepsnapshots}, and figure \ref{turbulentsnapshots} shows snapshots of the turbulent equilibrium state ($\epsilon = 16$). In these standard diagnostics, the distinction between the time-dependent and turbulent regimes is obscured: time series of TKE and individual structure snapshots do not reveal the key difference between them—namely, that the time-dependent regime exhibits temporal chaos while maintaining large-scale spatial coherence, whereas the turbulent regime is chaotic in both time and space. This transition is clearly revealed by examining a Hovmöller diagram of the statistical state in these regimes, as shown in figure \ref{Hovmollertimedepvsturb}.

\section{Comparison of solutions to observations}
\begin{figure}

\subfloat[]{%
            \includegraphics[width=.48\linewidth]{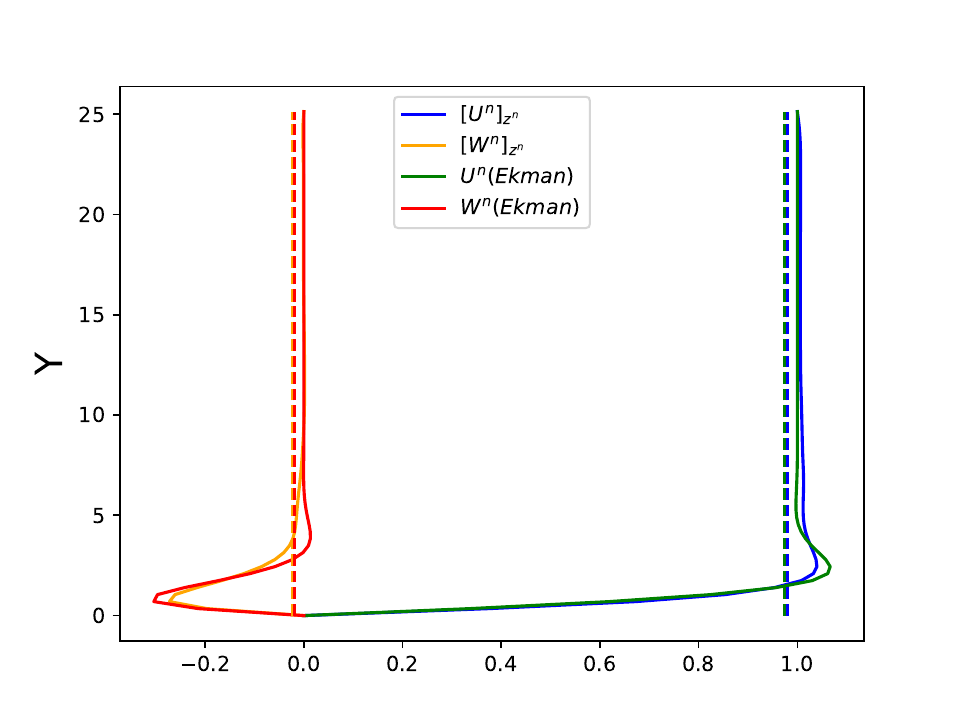}
            \label{subfig:a}%
        }\hfill
        \subfloat[]{%
            \includegraphics[width=.48\linewidth]{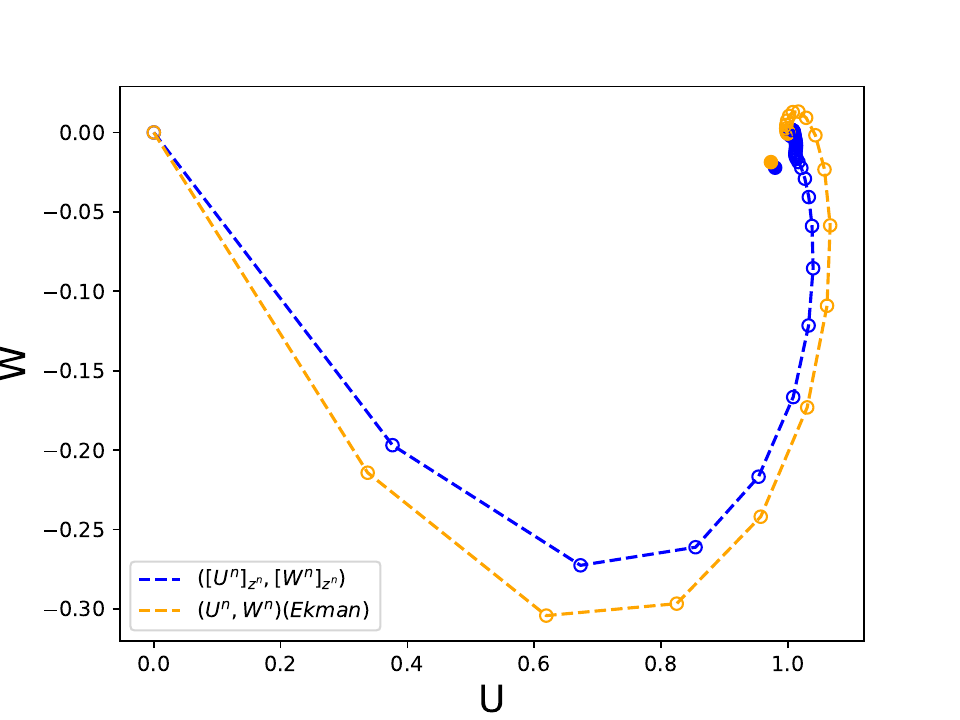}
            \label{subfig:b}%
        }\\
        \subfloat[]{%
            \includegraphics[width=.48\linewidth]{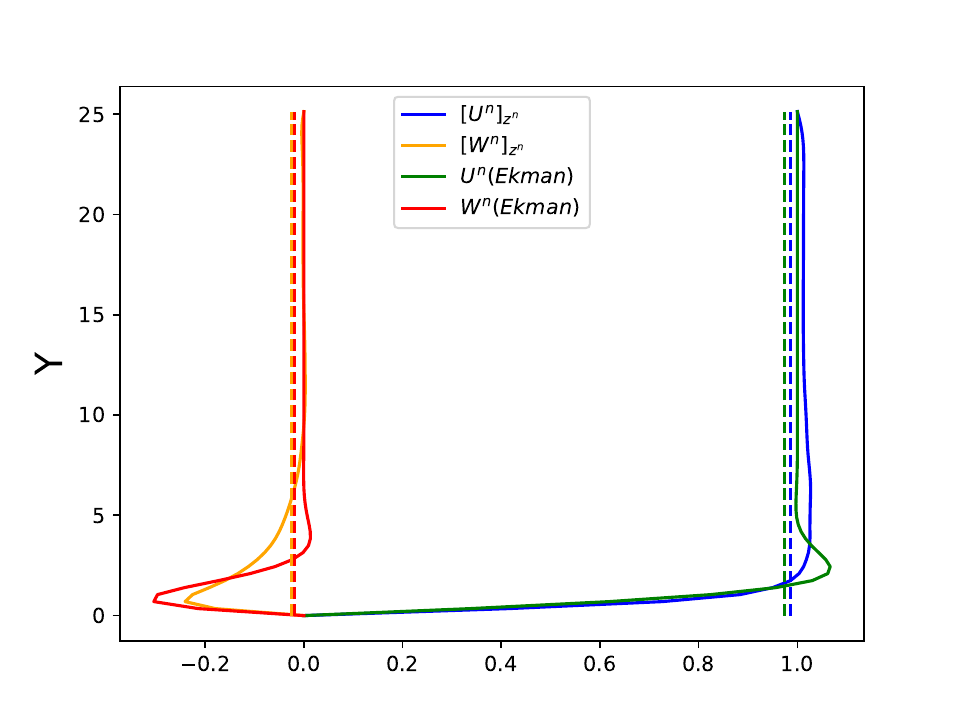}
            \label{subfig:c}%
        }\hfill
        \subfloat[]{%
            \includegraphics[width=.48\linewidth]{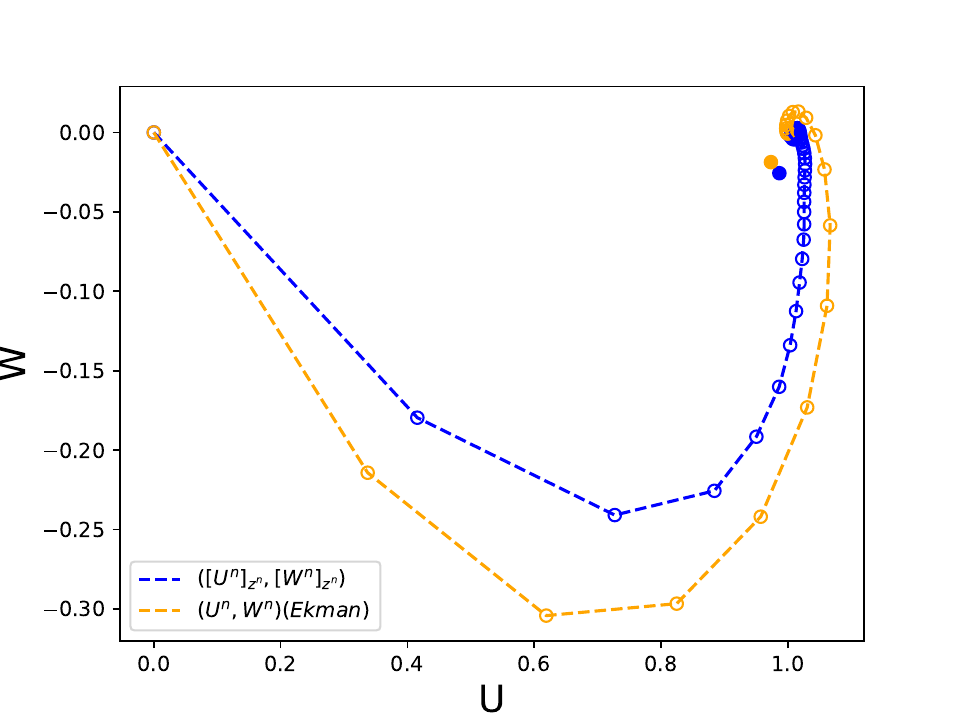}
            \label{subfig:d}%
        }\\
        \subfloat[]{%
            \includegraphics[width=.48\linewidth]{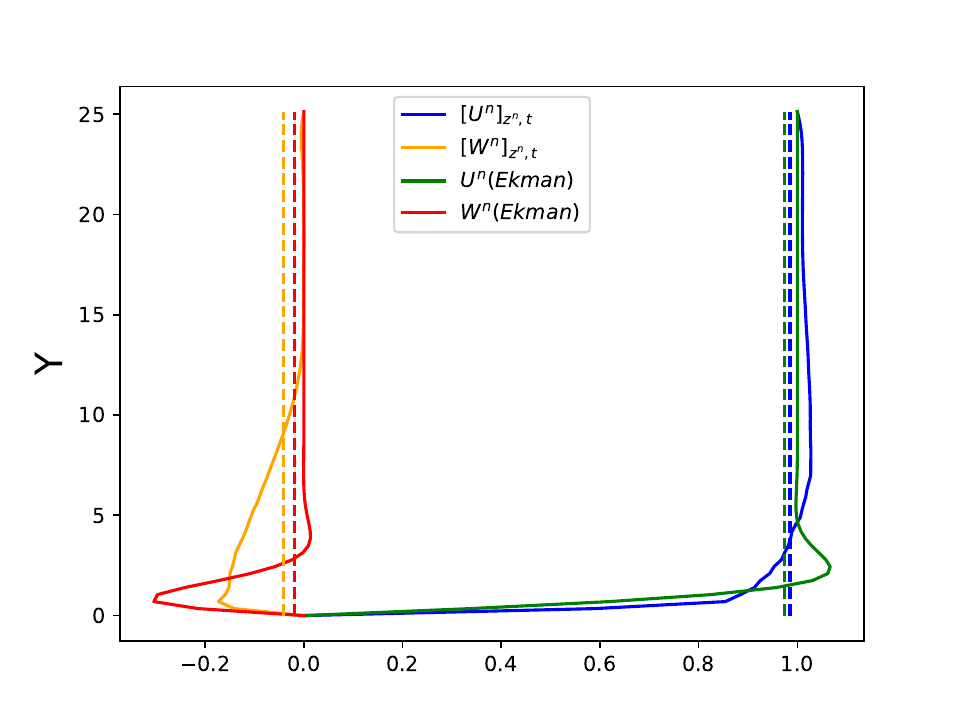}
            \label{subfig:e}%
        }\hfill
        \subfloat[]{%
            \includegraphics[width=.48\linewidth]{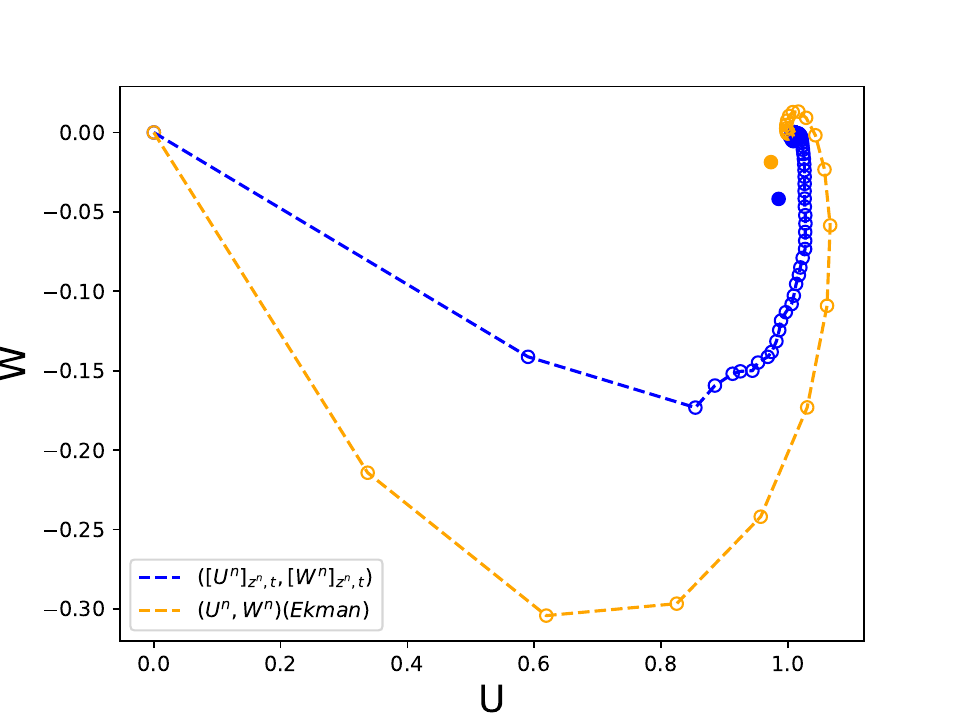}
            \label{subfig:f}%
        }
\caption{For the purpose of displaying the results shown in this figure the $15^o$ coordinate frame rotation has been undone so that the $x^{n}$ (indicated by a superscript n) is aligned with the geostrophic wind direction.   $[U^{n}]_{z^{n}}$ and $[W^{n}]_{z^{n}}$ as functions of $y$, along with a hodograph illustrating these velocities. Panels (a) and (b) correspond to $\epsilon = 0$, panels (c) and (d) correspond to $\epsilon = 2$, and panels (e) and (f) correspond to $\epsilon = 16$. }
\label{profilecomparisonvsepsilon}
\end{figure}

In order to make comparison with observed Ekman turbulent profiles  we undo the coordinate system rotation so that the coordinate system used for comparison with data is the unrotated coordinate system in which the variables are indicated by a superscript (n) e.g. coordinate $x^{n}$ is aligned with the geostrophic wind direction. A feature of primary dynamical significance in the PBL is the time, streamwise, and spanwise mean velocity structure of the Ekman state whether it be associated with a fixed point, a time-dependent state or a turbulent state.  Comparison is made between the analytic laminar solution for the Ekman velocity profile and the mean velocity structure obtained for fixed point and turbulent equilibria.
Shown in figure \ref{profilecomparisonvsepsilon} are the structures of $[U^{n}]_{z^{n}}$ and $[W^{n}]_{z^{n}}$ as functions of y and as a hodograph. Panels (a) and (b) correspond to $\epsilon=0$, for which equilibration to a fixed point occurs solely through the unstable inflectional mode. This demonstrates that a unique equilibrium Ekman profile compatible with the equations of motion is obtained by a small modification of the Ekman profile due to equilibration of the inflectional mode. We conclude that the observed collapsed Ekman profile cannot be produced by equilibration of the inflectional instability by its self-consistent induced fluxes.
Panels (c) and (d) correspond to $\epsilon=2$, which produces a weak field of background turbulence. The resulting equilibrium differs substantially from that produced by the inflectional mode alone and clearly shows the influence of the RS torque mechanism on the equilibration.
Panels (d) and (e), corresponding to $\epsilon=16$ (the turbulent regime), shows the influence on the equilibrated profile of stronger background turbulence support for the RS torque mechanism. Comparing the equilibrium profiles for $\epsilon=0$, $\epsilon=2$, and $\epsilon=16$, we infer that the observed collapse of the Ekman profile—reflected in the more horizontal orientation of the blue line in the hodograph for $\epsilon=16$—is generated by the RS torque mechanism.
The near-surface equilibrated profiles of $[U^{n}]_{z^{n}}$ and $[W^{n}]_{z^{n}}$ further indicate that, as $\epsilon$ increases, the surface wind becomes more closely aligned with the geostrophic wind direction than the $45^{\circ}$ veering predicted by the classical Ekman spiral. In summary, our equilibrated profiles for the strongly turbulent regime are consistent with previous simulations of the turbulent Ekman layer \citep{Coleman1992,Coleman1990,Coleman1999,Marlatt2012,Deusebio2014} and with observed PBL profiles \citep{Brown1970}, while finite amplitude equilibria produced by the inflectional mode are not.

\section{Implications of Equilibrium Regime Change for Transport in the PBL}

We have seen that the Ekman profile may equilibrate to a fixed point, or to a time-dependent state which may be a forced turbulence or a self-sustaining turbulence.  The dynamics of TC intensification is strongly influenced by radial transport which is controlled by the streamwise surface stress which in turn depends on the parameter $\epsilon$. 
To address the dependence of this intensification mechanism on transport in the outer core where Coriolis balance dominates, a local spanwise average of the mean streamwise and spanwise velocity equations are examined:\\
\begin{equation}
\partial_t[U]_{z}=-\frac{\partial [UV]_{z}}{\partial y}-\frac{\partial [u'v']_{x^{},z^{}}}{\partial y}+\frac{1}{Re}\frac{\partial^2 [U^{}]_{z^{}}}{\partial y^2} -\frac{2}{Re}[W^{}]_{z^{}}+\frac{2}{Re}sin(\theta)
\label{zavgUeq}
\end{equation}
\begin{equation}
\partial_t[W]_z=-\frac{d[WV]_z}{dy}-\frac{d[w'v']_{x,z}}{dy}+\frac{1}{Re}\frac{d^2 [W]_z}{dy^2} +\frac{2}{Re}[U]_z-\frac{2}{Re}cos(\theta),
\label{zavgWeq}
\end{equation}
in which the terms $\frac{2}{Re}sin(\theta)$ and $-\frac{2}{Re}cos(\theta)$  arise from the background pressure gradient. Explicitly,
$-[\frac{\partial P}{\partial x}]_z=\frac{2}{Re}sin(\theta)$ 
and $-[\frac{\partial P}{\partial z}]_z=-\frac{2}{Re}cos(\theta)$\\

In steady state, the left-hand side of equations  \eqref{zavgUeq} and \eqref{zavgWeq}  vanish:\\

\begin{equation}
0=-\frac{d [UV]_{z}}{d y}-\frac{d [u'v']_{x,z}}{d y}+\frac{1}{Re}\frac{d^2 [U]_{z}}{dy^2} -\frac{2}{Re}[W]_{z}+\frac{2}{Re}sin(\theta)
\label{steadyzavgUeq}
\end{equation}
\begin{equation}
0=-\frac{d[WV]_z}{dy}-\frac{d[w'v']_{x,z}}{dy}+\frac{1}{Re}\frac{d^2 [W]_z}{dy^2} +\frac{2}{Re}[U]_z-\frac{2}{Re}cos(\theta)
\label{steadyzavgWeq}
\end{equation}\\

Taking the y average of  equations \eqref{steadyzavgUeq}   \label{steadyzavgWeq} yields:\\

\begin{equation}
0=-\frac{1}{Re \cdot Ly}\frac{d [U]_z}{dy}(y=0) -\frac{2}{Re}[W]_{y,z}+\frac{2}{Re}sin(\theta)
\label{yzavgUeq}
\end{equation}
\begin{equation}
0=-\frac{1}{Re \cdot Ly}\frac{d [W]_z}{dy}(y=0)+\frac{2}{Re}[U]_{y,z}-\frac{2}{Re}cos(\theta)
\label{yzavgWeq}
\end{equation}\\

Rearranging equations \eqref{yzavgUeq} and \eqref{yzavgWeq} to write $[U]_{y,z}$ and $[W]_{y,z}$ in terms of surface stress:
\begin{equation}
[W]_{y,z}=-\frac{1}{2 \cdot Ly}\frac{d [U]_z}{dy}(y=0)+sin(\theta)
\label{Usurfstresseq}
\end{equation}
\begin{equation}
[U]_{y,z}=\frac{1}{ 2\cdot Ly}\frac{d [W]_z}{dy}(y=0)+cos(\theta)
\label{Wsurfstresseq}
\end{equation}\\
In order to make comparison with data we now transform \eqref{Usurfstresseq} and \eqref{Wsurfstresseq} to the unrotated coordinate system  $[U^{n}]_{z^{n}}$ and $[W^{n}]_{z^{n}}$.  This transformation  is shown explicitly in appendix \ref{appendix:coordinate} and yields for the convergent bulk transport flux:\\

\begin{equation}
[W^{n}]_{y,z^{n}}=-\frac{1}{2 \cdot Ly}\frac{d[U^{n}]_{z^{n}}}{dy}(y=0).
\label{Winteneq}
\end{equation}\\

In wall-bounded shear flow turbulence, increasing \(\epsilon\) leads to a regime with enhanced wall stress, \(-\frac{1}{Re}\frac{d [U^{n}]_{z^{n}}}{dy}\big|_{y=0}\) \citep{Farrell-Ioannou-2012}. Consequently, the first term on the right-hand side of equation~\eqref{Winteneq} becomes more negative, yielding more negative values of \([W^{n}]_{y,z^{n}}\). In the case of a TC, this would influence the deepening process by enhancing transport into the low-pressure center from the outer core.

In the Ekman equilibrium state, surface stresses control the transport of both momentum and tracers such as potential temperature. In addition, the vertical distribution of radial velocity plays a key role in determining tracer transport. These dynamically important transports vary with the equilibrium regime, as illustrated in figure~\ref{profilecomparisonvsepsilon}. The competition between momentum-transport mediated dissipation and intensification driven by the convergence of high–potential-temperature air from the outer core into the low-pressure center plays a crucial role in the TC deepening process. This balance can shift abruptly as turbulence intensity varies, such as during a transition from a fixed-point regime to a turbulent RSS regime.  The inflow depth appears to be particularly sensitive to abrupt increase with transition to a turbulent Ekman regime as indicated in ~\ref{profilecomparisonvsepsilon} panel $(e)$.   We note that this increase in inflow depth is consistent with the requirement of equilibrium that the curvature not change sign in the region above the viscous boundary while the bulk transport equals that required by the surface boundary stress.  Although gradient-wind balance, rather than Coriolis balance, governs the inner core of a TC, similar regime transitions are expected.
\section{RSS maintenance mechanism}
\label{RSSmaintainsection}
We now examine the mechanisms that maintain the RSS in turbulent Ekman dynamics and their relationship to the mechanisms sustaining the RSS in wall-bounded shear flows \citep{Farrell2016}. From this point on, we once again adopt the rotated coordinate system.
The mechanism governing RSS dynamics and equilibration can be analyzed through the balance equations for the RSS streak velocity, $U_s = U - [U]_z$, and the roll vorticity, $\mathcal{Z}_r = \mathcal{Z} - [\mathcal{Z}]_z$, where the streamwise vorticity component is defined as $\mathcal{Z} = -\partial V/\partial z + \partial W/\partial y$. The balance of the streak component, $U_s$, is described by the following equation:\\
\begin{equation}
\partial_t U_{s}=-(\partial_y(UV)-\partial_y[UV]_z)-\partial_z(UW)-(\partial_y[u'v']_x-\partial_y[u'v']_{x,z})-\partial_z[u'w']_x+\Delta_1 \frac{U_{s}}{Re}-\frac{2}{Re}(W-[W]_z),
\end{equation}\\

Given that streaks of both signs occur, in order to obtain a linear measure of streak forcing, each term is multiplied by $sign(U_{s})$.  The equations for the physically distinct terms in the dynamics maintaining $U_{s}$ are:\\
\begin{equation}I_A=sign(U_{s})\cdot(-(V\frac{\partial U}{\partial y}-[V\frac{\partial U}{\partial y}]_z)-(W\frac{\partial U}{\partial z}-[W\frac{\partial U}{\partial z}]_z))\end{equation}
\begin{equation}
I_B=sign(U_{s})\cdot(-([v'\frac{\partial u'}{\partial y}]_x-[v'\frac{\partial u'}{\partial y}]_{x,z})-([w'\frac{\partial u'}{\partial z}]_x-[w'\frac{\partial u'}{\partial z}]_{x,z}))
\end{equation}
\begin{equation}
I_C=sign(U_{s})\cdot(\frac{1}{Re}\Delta_1 U_{st})
\end{equation}
\begin{equation}
I_D=sign(U_s) \cdot (-\frac{2}{Re}(W-[W]_z))
\end{equation}\\

These components of streak forcing are identified as lift-up ($I_A$), fluctuation Reynolds stress ($I_B$), viscous damping ($I_C$), and the Coriolis force ($I_D$).\\

The balance equation for the roll vorticity is:\\ 

\begin{equation}
\begin{split}
\partial_t\mathcal{Z}_r=-(V_r\partial_y\mathcal{Z}_r+(W_r+[W]_z)\partial_z\mathcal{Z}_r-[V_r \partial_y\mathcal{Z}_r]_z-[W_r\partial_z \mathcal{Z}_r]_z)- V_r\partial_{yy}[W]_z\\+(\partial_{zz}-\partial_{yy})(\overline{v'w'})-\partial_{yz}(\overline{w'^2}-\overline{v'^2})+\Delta_1\frac{\mathcal{Z}_r}{Re}+\frac{2}{Re}\partial_y U_s
\end{split}
\end{equation}\\

The equations for the physically distinct terms in the dynamics maintaining $\partial_t\mathcal{Z}_r$ are:\\

\begin{equation}
I_E=sign(\mathcal{Z}_r) \cdot (-(V_r\partial_y\mathcal{Z}_r+(W_r+[W]_z)\partial_z\mathcal{Z}_r-[V_r \partial_y\mathcal{Z}_r]_z-[W_r\partial_z \mathcal{Z}_r]_z))
\end{equation}
\begin{equation}
I_F=sign(\mathcal{Z}_r)\cdot (-V_r\partial_{yy}[W]_z)
\end{equation}
\begin{equation}
I_G=sign(\mathcal{Z}_r) \cdot ((\partial_{zz}-\partial_{yy})(\overline{v'w'})-\partial_{yz}(\overline{w'^2}-\overline{v'^2}))\end{equation}

\begin{equation}
I_H= sign(\mathcal{Z}_r) \cdot \Delta_1\frac{\mathcal{Z}_r}{Re}
\end{equation}
\begin{equation}
I_I=sign(\mathcal{Z}_r) \cdot \frac{2}{Re}\partial_y U_s
\end{equation}\\

The terms $I_E, I_F, I_G, I_H,I_I$ represent the individual contributions to roll vorticity production. Specifically, $I_E$ arises from advection of roll vorticity, $I_F$ arises from interaction with the background streamwise vorticity gradient ($\partial_y[\mathcal{Z}]_z = \partial_{yy}[W]_z$); $I_G$ from Reynolds stress divergence; $I_H$ from diffusion; and $I_I$ from the tilting of planetary vorticity.
\begin{figure}
\centering{
\begin{subfigure}{0.8\textwidth}\caption{}
\includegraphics[width=\linewidth]{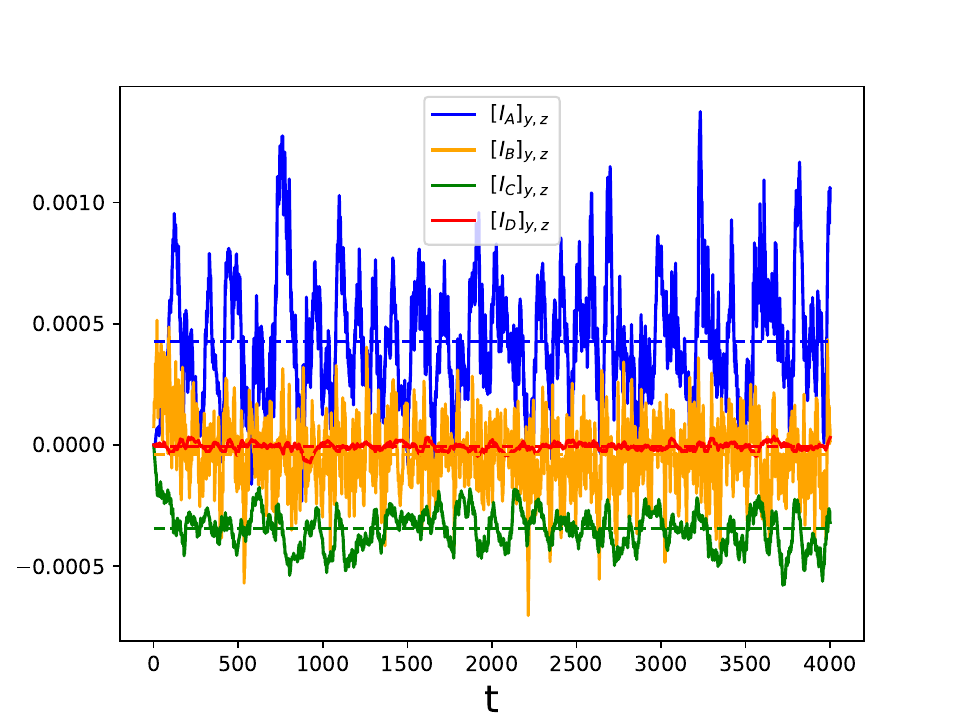}
\end{subfigure}
\begin{subfigure}{0.8\textwidth}\caption{}
\includegraphics[width=\linewidth]{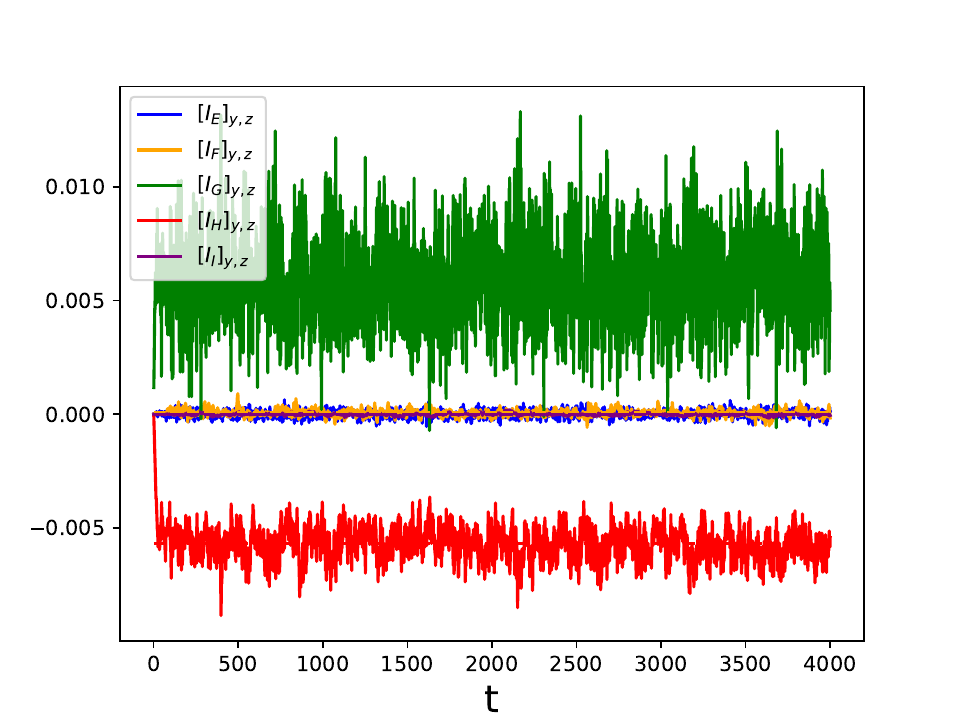}
\end{subfigure}
}
\caption{
Equilibrium balance of the forcing in time-dependent RSS turbulence, averaged over the y and z directions. Panel (a) shows the maintenance of the streak $U_s$, decomposed into the lift-up contribution $I_A$, the Reynolds-stress contribution $I_B$, the diffusive contribution $I_C$, and the Coriolis contribution $I_D$. Panel (b) shows the maintenance of the roll vorticity $\zeta_r$, decomposed into advection contribution $I_E$, background streamwise vorticity gradient contribution $I_F$, the Reynolds-stress contribution $I_G$, the diffusive contribution $I_H$, and the Coriolis contribution $I_I$. 
}
\label{fig:streakrollmaintenanceturbulence}
\end{figure}
Time series of the components maintaining the streak and roll in PBL RSS turbulence are shown in Figure \ref{fig:streakrollmaintenanceturbulence}. The streak maintenance mechanism in PBL RSS turbulence is very similar to that of wall-bounded shear flow turbulence in which forcing of the streak by lift-up balances transfer of energy from the streak by the resolved fluctuation Reynolds stresses, and viscous dissipation.  Transfer of energy from the streak by the resolved Reynolds stresses serves both to maintain the fluctuations and to exert feedback regulation on the streak amplitude,

The mechanism maintaining roll vorticity, $\mathcal{Z}_r$, is also the same as that maintaining roll vorticity in wall-bounded shear flow turbulence in which the Reynolds stress vorticity source term balances dissipation. 
The inflectional instability mechanism, which arises from $\frac{\partial^2 [W]_z}{\partial y^2}$ \citep{Lilly1966} and is contained in $I_{F}$, is shown by our results to be unrelated to the equilibrium maintenance of roll vorticity $\mathcal{Z}_{r}$ in the PBL. Instead, $\mathcal{Z}_{r}$ at equilibrium is sustained against dissipation, $I_{H}$, by vorticity forcing associated with the fluctuation Reynolds stress term, $I_{G}$.
In summary, analysis of the mechanisms sustaining the streak velocity, $U_s$, and the roll vorticity, $\mathcal{Z}_r$  at equilibrium in the PBL, shows that the processes maintaining $U_s$ and $\mathcal{Z}_r$ in PBL RSS turbulence are those operating in wall-bounded shear flows. Furthermore, the inflectional instability mechanism associated with $\frac{\partial^2 [W]_z}{\partial y^2}$ does not contribute to RSS maintenance. We note that the negligible contribution of the Coriolis force term to maintaining both $U_s$ and $\mathcal{Z}_{r}$ aligns with previous findings \citep{Gao2014,Gao2015}.

\section{Fluctuation Vorticity Maintenance Mechanism}
\label{fluctuationmaintainsection}
We now examine the mechanisms sustaining enstrophy at turbulent equilibrium. The total vorticity vector, \(\boldsymbol{\omega}\), with components in the \(x\), \(y\), and \(z\) directions, is denoted by \((\zeta, \eta, \chi)\). The total vorticity  can be decomposed into a streamwise mean component and a fluctuating component:\\

\begin{equation}
\bv{\omega}(x,y,z,t)=\bv{\Theta}(y,z,t)+\bv{\omega}'(x,y,z,t)
\end{equation}
\begin{equation}
\overline{\bv{\omega}}=\bv{\Theta}
\end{equation}
\begin{equation}
\zeta(x,y,z,t)=\mathcal{Z}(y,z,t)+\zeta'(x,y,z,t)
\end{equation}
\begin{equation}\eta(x,y,z,t)= \mathcal{H}(y,z,t)+\eta'(x,y,z,t)\end{equation}
\begin{equation}\chi(x,y,z,t)=\mathcal{X}(y,z,t)+\chi'(x,y,z,t)
\end{equation}
\begin{equation}
\overline{(\zeta,\eta,\chi)}
=(\mathcal{Z},\mathcal{H}, \mathcal{X})
\end{equation}\\

We begin by examining the mechanisms sustaining the fluctuating streamwise vorticity, denoted by $\zeta'$. The governing equation for the balance of the fluctuating streamwise vorticity component is:\\
\begin{equation}
\partial_t \zeta'=-(\bv{U}\cdot \nabla)\zeta'- (\bv{u'}\cdot \nabla)([\mathcal{Z}]_z+(\mathcal{Z}-[\mathcal{Z}]_z)+(\bv{\Theta}\cdot \nabla){u}'+(\bv{\omega'}\cdot \nabla)U+\frac{\Delta \zeta'}{Re}+\frac{2}{Re}\frac{\partial u'}{\partial y}
\end{equation}\\

The equation for the enstrophy of the fluctuating streamwise vorticity component is obtained by multiplying by $\zeta'$ and then taking the mean:\\

\begin{equation}
\partial_t \frac{\overline{\zeta'^2}}{2}=-(\bv{U}\cdot \nabla)\frac{\overline{\zeta'^2}}{2}- (\overline{\zeta'\bv{u'}}\cdot \nabla)([\mathcal{Z}]_z+(\mathcal{Z}-[\mathcal{Z}]_z)+(\bv{\Theta}\cdot \overline{(\nabla{u}')\zeta'})+(\overline{\zeta'\bv{\omega'}}\cdot \nabla)U-\frac{\overline{\nabla \zeta'\cdot \nabla \zeta'}}{Re}+\frac{2}{Re}\overline{\zeta'\frac{\partial u'}{\partial y}}
\end{equation}
\begin{equation}
I_{\zeta'A}=-(\bv{U}\cdot \nabla)\frac{\overline{\zeta'^2}}{2}
\end{equation}
\begin{equation}
I_{\zeta'B}=- (\overline{\zeta'\bv{u'}}\cdot \nabla)[\mathcal{Z}]_z=-\overline{\zeta'v'}\frac{\partial^2 [W]_z }{\partial y^2}
\end{equation}
\begin{equation}
I_{\zeta'C}=- (\overline{\zeta'\bv{u'}}\cdot \nabla)(\mathcal{Z}-[\mathcal{Z}]_z)
\end{equation}
\begin{equation}
I_{\zeta'D}=(\bv{\Theta}\cdot \overline{(\nabla{u}')\zeta'})
\end{equation}
\begin{equation}
I_{\zeta'E}=(\overline{\zeta'\bv{\omega'}}\cdot \nabla)U
\end{equation}
\begin{equation}
I_{\zeta'F}=-\frac{\overline{\nabla \zeta'\cdot \nabla \zeta'}}{Re}
\end{equation}
\begin{equation}
I_{\zeta' G}=\frac{2}{Re}\overline{\zeta'\frac{\partial u'}{\partial y}}
\end{equation}\\
\begin{figure}
\centering{
\includegraphics[width=\linewidth]{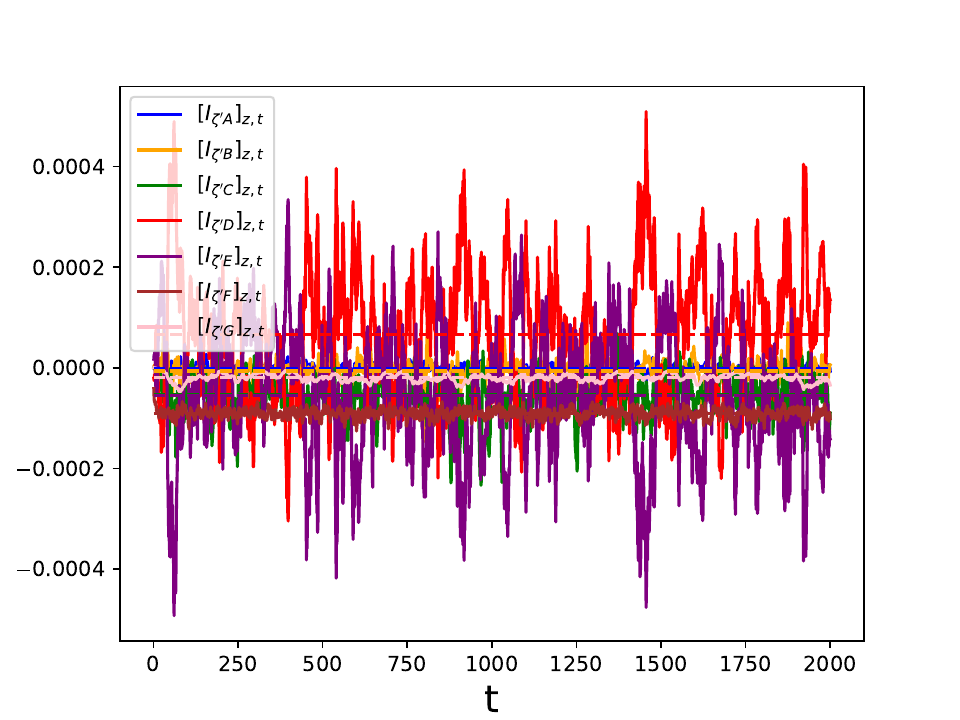}}
\caption{Components of the balance governing the streamwise fluctuation enstrophy, $\overline{\zeta'^2}/2$, in turbulent equilibrium for parameters $\epsilon = 16$ and $Re = 300$. All terms are averaged in both the spanwise direction $z$ and the wall-normal direction $y$. }
\label{omegaxmaintenance}
\end{figure}

The contributions to the fluctuation streamwise enstrophy forcing are categorized as follows: advection, $(I_{\zeta'A})$; production of fluctuation streamwise enstrophy by $\frac{\partial^2 [W]_z}{\partial y^2}$, $(I_{\zeta'B})$; production by the remaining components of the mean vorticity, $(I_{\zeta'C})$; stretching and tilting of fluctuation streamwise vorticity by fluctuation strain, $(I_{\zeta'D})$; stretching and tilting of fluctuation streamwise vorticity by mean strain, $(I_{\zeta'E})$; viscous dissipation, $(I_{\zeta'F})$; and stretching and tilting of planetary vorticity, $(I_{\zeta'G})$. 

The time series of the spatially averaged diagnostics $I_{\zeta'A}, I_{\zeta'B}, I_{\zeta'C}, I_{\zeta'D}, I_{\zeta'E}, I_{\zeta'F}, I_{\zeta'G}$ is shown in figure \ref{omegaxmaintenance}. The leading balance comprises a positive contribution from stretching by the fluctuating strain, $I_{\zeta'D}$, and a negative contribution from dissipation, $I_{\zeta'F}$.

We now examine the mechanism responsible for maintaining the fluctuating wall-normal vorticity component. The governing equation for the balance of this fluctuating wall-normal vorticity component is:

\begin{equation}
\partial_t \eta'=-(\bv{U}\cdot \nabla)\eta'- (\bv{u'}\cdot \nabla)\mathcal{H}+(\bv{\Theta}\cdot \nabla){v}'+(\bv{\omega'}\cdot \nabla)V+\frac{\Delta \eta'}{Re}+\frac{2}{Re}\frac{\partial v'}{\partial y}
\end{equation}
The equation for the enstrophy associated with the fluctuating wall-normal vorticity component is obtained by multiplying by $\eta'$ and then taking a streamwise average.\\

\begin{equation}
\partial_t \frac{\overline{\eta'^2}}{2}=-(\bv{U}\cdot \nabla)\overline{\eta'^2}- (\overline{\eta'\bv{u'}}\cdot \nabla)\mathcal{H}+(\bv{\Theta}\cdot \overline{(\nabla{v}')\eta'})+(\overline{\eta'\bv{\omega'}}\cdot \nabla)V-\frac{\overline{\nabla \eta'\cdot \nabla \eta'}}{Re}+\frac{2}{Re}\overline{\eta'\frac{\partial v'}{\partial y}}
\end{equation}
\begin{equation}
I_{\eta'A}=-(\bv{U}\cdot \nabla)\overline{\eta'^2}
\end{equation}
\begin{equation}
I_{\eta'C}=- (\overline{\eta'\bv{u'}}\cdot \nabla)\mathcal{H}
\end{equation}
\begin{equation}
I_{\eta' D}=(\bv{\Theta}\cdot \overline{(\nabla{v}')\eta'})
\end{equation}
\begin{equation}
I_{\eta' E}=(\overline{\eta'\bv{\omega'}}\cdot \nabla)V
\end{equation}
\begin{equation}
I_{\eta'F}=-\frac{\overline{\nabla \eta'\cdot \nabla \eta'}}{Re}
\end{equation}
\begin{equation}
I_{\eta'G}=\frac{2}{Re}\overline{\eta'\frac{\partial v'}{\partial y}}
\end{equation}\\
\begin{figure}
\centering{
\includegraphics[width=\linewidth]{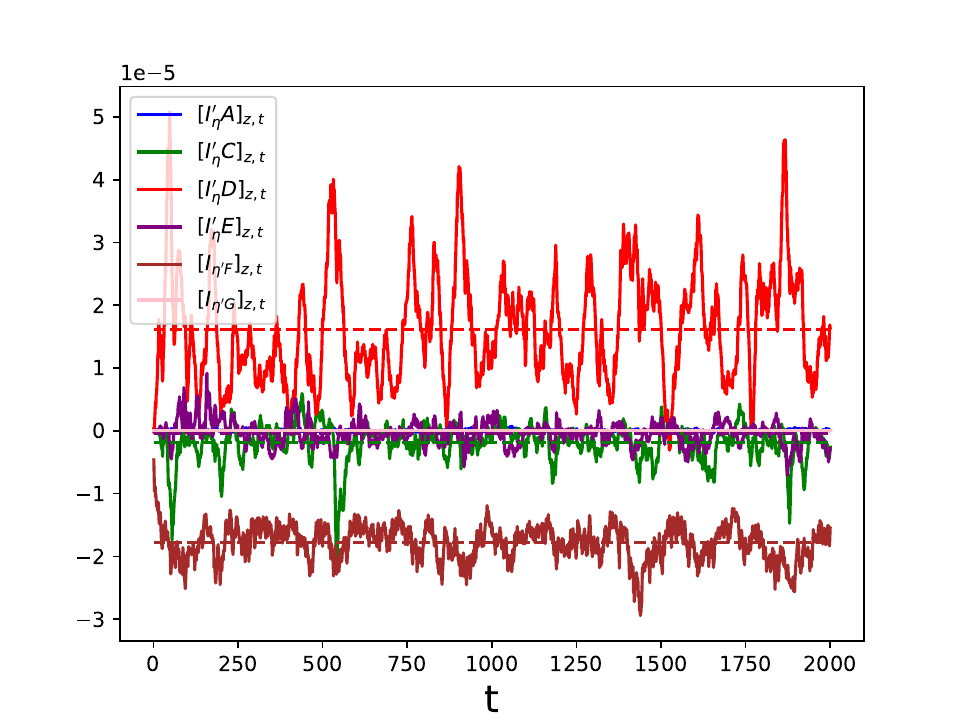}}
\caption{Components of the balance sustaining the wall-normal fluctuation enstrophy, $\overline{\eta'^2}/2$, in turbulent equilibrium for parameters $\epsilon = 16$ and $Re = 300$. The terms shown are averaged in both the spanwise (z) and wall-normal (y) directions.}
\label{omegaymaintenance}
\end{figure}

The components of the fluctuation wall-normal enstrophy forcing are classified as follows: advection, $(I_{\eta'A})$; fluctuation wall-normal enstrophy production due to mean wall-normal vorticity, $(I_{\eta'C})$; stretching/tilting of fluctuation wall-normal vorticity by fluctuation strain, $(I_{\eta'D})$; stretching/tilting of fluctuation wall-normal vorticity by mean strain, $(I_{\eta'E})$; viscous damping, $(I_{\eta'F})$; and stretching/tilting of planetary vorticity, $(I_{\eta'G})$.
The time series of the spatially averaged diagnostics $I_{\eta'A}I_{\eta'C}, I_{\eta'D}, I_{\eta'E}, I_{\eta'F}, I_{\eta'G}$ is shown in figure \ref{omegaymaintenance}. The dominant balance consists of a positive contributions from stretching by the fluctuating strain, $I_{\eta'D}$, and a negative contributions from dissipation, $I_{\zeta'F}$. 

Lastly, we examine the mechanism responsible for sustaining the fluctuating spanwise vorticity component. The governing equation for the balance of this fluctuating spanwise vorticity is: 

\begin{equation}
\partial_t \chi'=-(\bv{U}\cdot \nabla)\chi'- (\bv{u'}\cdot \nabla)([\mathcal{X}]_z+(\mathcal{X}-[\mathcal{X}]_z))+(\bv{\Theta}\cdot \nabla){w}'+(\bv{\omega'}\cdot \nabla)W+\frac{\Delta \chi'}{Re}+\frac{2}{Re}\frac{\partial w'}{\partial y}
\end{equation}
Equations for enstrophy associated with the fluctuation spanwise vorticity component is obtained by multiplying $\chi'$ and taking streamwise mean

\begin{equation}
\partial_t \frac{\overline{\chi'^2}}{2}=-(\bv{U}\cdot \nabla)\frac{\overline{\chi'^2}}{2}- (\overline{\chi'\bv{u'}}\cdot \nabla)([\mathcal{X}]_z+(\mathcal{X}-[\mathcal{X}]_z)+(\bv{\Theta}\cdot \overline{(\nabla{w}')\chi'})+(\overline{\chi'\bv{\omega'}}\cdot \nabla)W-\frac{\overline{\nabla \chi'\cdot \nabla \chi'}}{Re}+\frac{2}{Re}\overline{\chi'\frac{\partial w'}{\partial y}}
\end{equation}
\begin{equation}
I_{\chi'A}=-(\bv{U}\cdot \nabla)\frac{\overline{\chi'^2}}{2}
\end{equation}
\begin{equation}I_{\chi' B}=
- (\overline{\chi'\bv{u'}}\cdot \nabla)[\mathcal{X}]_z=\overline{\chi'v'}\frac{\partial^2 [U]_z}{\partial y^2}
\end{equation}
\begin{equation}
I_{\chi' C}=
- (\overline{\chi'\bv{u'}}\cdot \nabla)(\mathcal{X}-[\mathcal{X}]_z)
\end{equation}
\begin{equation}
I_{\chi' D}=(\bv{\Theta}\cdot \overline{(\nabla{w}')\chi'})
\end{equation}
\begin{equation}
I_{\chi' E}=(\overline{\chi'\bv{\omega'}}\cdot \nabla)W
\end{equation}
\begin{equation}
I_{\chi'F}=-\frac{\overline{\nabla \chi'\cdot \nabla \chi'}}{Re}
\end{equation}
\begin{equation}
I_{\chi' G}=\frac{2}{Re}\overline{\chi'\frac{\partial w'}{\partial y}}
\end{equation}\\
\begin{figure}
\centering{
\includegraphics[width=\linewidth]{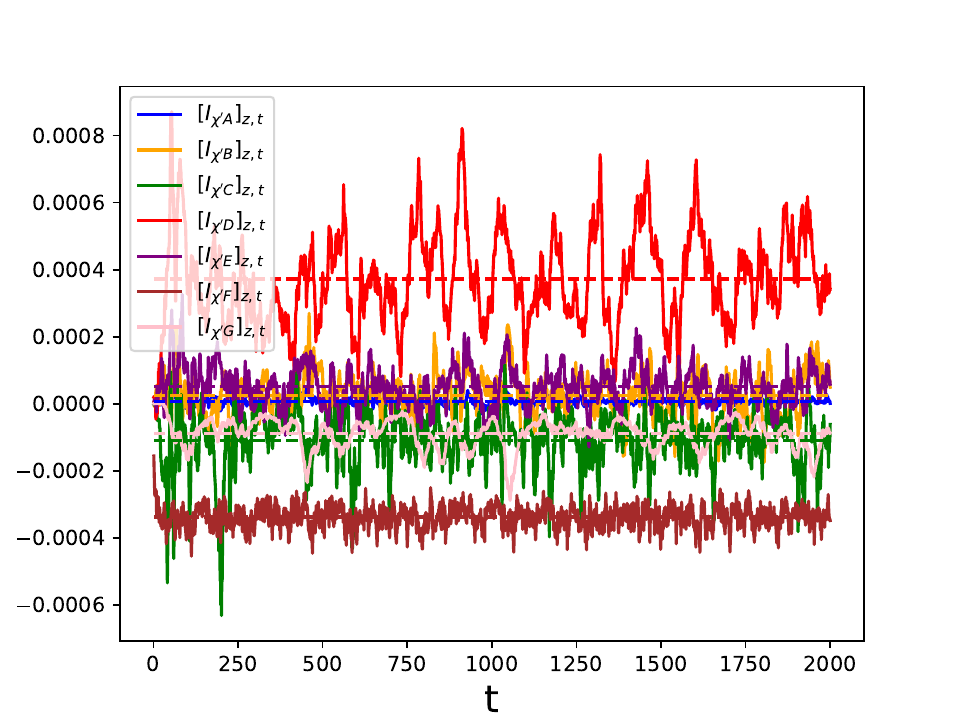}}
\caption{Components of the balance maintaining the fluctuation spanwise enstrophy $\overline{\chi'^2}/2$ in the turbulent equilibrium for parameters $\epsilon=16$; $Re = 300$. Terms shown are averaged in both the spanwise direction, z, and the wall normal direction, y}
\label{omegazmaintenance}
\end{figure}

The components of the fluctuation spanwise enstrophy forcing are classified as follows: advection, $(I_{\chi'A})$; fluctuation spanwise enstrophy production by $-\frac{\partial^2 [U]_z}{\partial y^2}$, $(I_{\chi'B})$; fluctuation spanwise enstrophy production by the remaining components of the mean spanwise vorticity, $(I_{\chi'C})$; stretching/tilting of fluctuation spanwise vorticity by fluctuation strain, $(I_{\chi'D})$; stretching/tilting of fluctuation spanwise vorticity by mean strain, $(I_{\chi'E})$; viscous damping, $(I_{\chi'F})$; and stretching/tilting of planetary vorticity, $(I_{\chi'G})$. 

The time series of the spatially averaged diagnostics $I_{\chi'A}, I_{\chi'B}, I_{\chi'C}, I_{\chi'D}, I_{\chi'E}, I_{\chi'F}, I_{\chi'G}$ is shown in figure~\ref{omegazmaintenance}. The dominant balance is between a positive contributions from stretching by fluctuation strain, $I_{\chi'D}$,  and the negative contributions from diffusion, $I_{\chi'F}$.

In summary, our analysis of the mechanisms sustaining all three components of the fluctuation vorticity, $\boldsymbol{\omega}' = (\zeta', \eta', \chi')$, shows that each component is maintained primarily by stretching and tilting of fluctuation vorticity by the fluctuation strain, $I_{\zeta'D}, I_{\eta'D}, I_{\chi'D}$, in opposition to dissipation, $I_{\zeta'F}, I_{\eta'F}, I_{\chi'F}$. This behavior is consistent with a strongly operating self-sustaining turbulence mechanism \citep{Farrell-Ioannou-2012}.

Furthermore, the results indicate that the vorticity gradients associated with $\frac{\partial^2[W]_z}{\partial y^2}$ and $-\frac{\partial ^2[U]_z}{\partial y^2}$ do not contribute to the maintenance of the fluctuation streamwise or spanwise vorticity. This, in turn, implies that the inflectional instabilities related to $\frac{\partial^2[W]_z}{\partial y^2}$ and $-\frac{\partial ^2[U]_z}{\partial y^2}$ do not play a role in sustaining the fluctuation vorticity, $\boldsymbol{\omega}' = (\zeta', \eta', \chi')$, in the turbulent regime.
From the preceding section \ref{RSSmaintainsection} and the present section \ref{fluctuationmaintainsection}, we conclude that turbulent equilibria involving both the 
roll vorticity in the mean flow, $\mathbf{U}$, and the fluctuation covariance, $\mathbf{C}$, are maintained primarily by the Reynolds-stress torque mechanism \citep{Farrell-Ioannou-2012,Farrell2016}. Moreover, the vorticity gradients associated with the inflectional instability mechanism, $\partial^2 [U]_{z}/\partial y^2$ and $\partial^2 [W]_{z}/\partial y^2$, make negligible contributions to sustaining either the mean flow, $\mathbf{U}$, or the fluctuation covariance, $\mathbf{C}$.
Observational studies of the atmospheric boundary layer show that the measured turbulence intensity—quantified by the RMS velocity fluctuations—typically lies between about 10\% and 30\% of the geostrophic wind speed.\citep{Friedrich2012, Watkins2012, Duncan2019, Barthelmie2014} This range corresponds to a strongly turbulent regime in our equilibrium regime diagram. We therefore infer that inflectional instability is not the primary mechanism governing the dynamics of the observed PBL, as the turbulent regime is predominantly maintained by the RS torque mechanism.

\section{The Self-Sustaining Process in PBL Turbulence}

We have seen that the mechanism of turbulence in the PBL is essentially the same as that in wall-bounded shear flows. Wall-bounded shear flows  maintain turbulence even after the initial excitation that triggered the transition to a turbulent state has been removed—a phenomenon known as transition to turbulence. It is therefore reasonable to ask whether PBL RSS turbulence is likewise self-sustaining. This can be tested diagnostically by setting $\epsilon = 0$ once the turbulent state has been established.
When $\epsilon=0$, the S3T equations reduce to: 
\begin{equation}
\frac{\partial \bv \Gamma}{\partial t}=\mathbf G(\bv \Gamma) + \sum_k \mathbf L_{RS}\mathbf C_k
\end{equation}
\begin{equation}
\frac{\partial \mathbf C_k}{\partial t}=\mathbf A \mathbf C_k+\mathbf C_k \mathbf A^{\dagger}
\end{equation}

\begin{figure}
\subfloat[]{%
            \includegraphics[width=.48\linewidth]{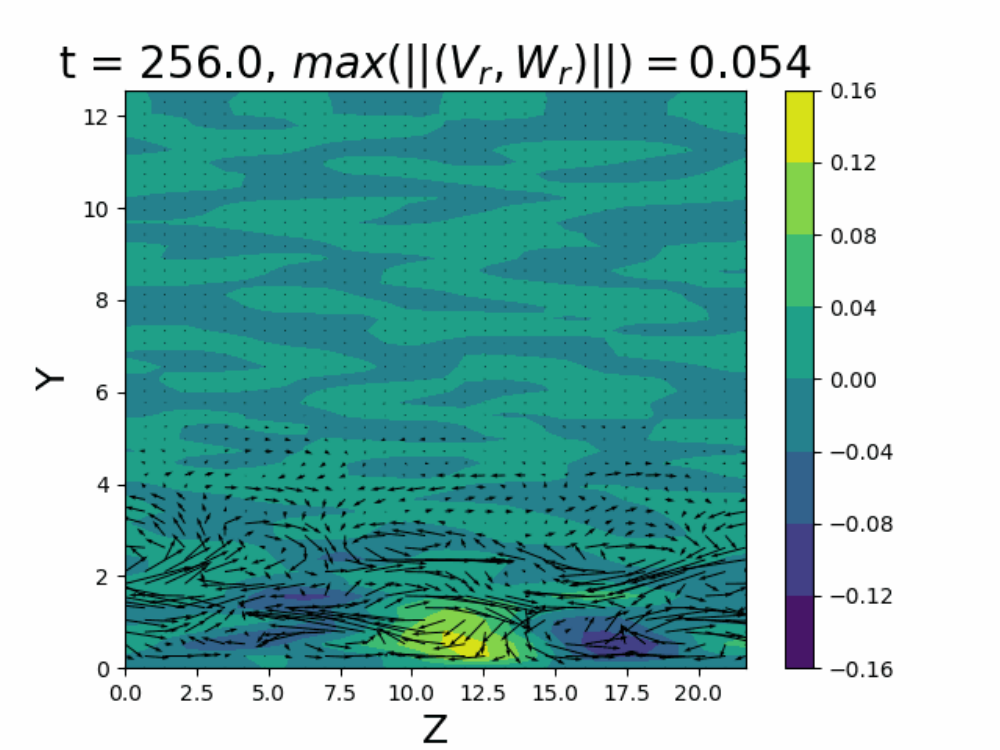}
            \label{subfig:a}%
        }\hfill
        \subfloat[]{%
            \includegraphics[width=.48\linewidth]{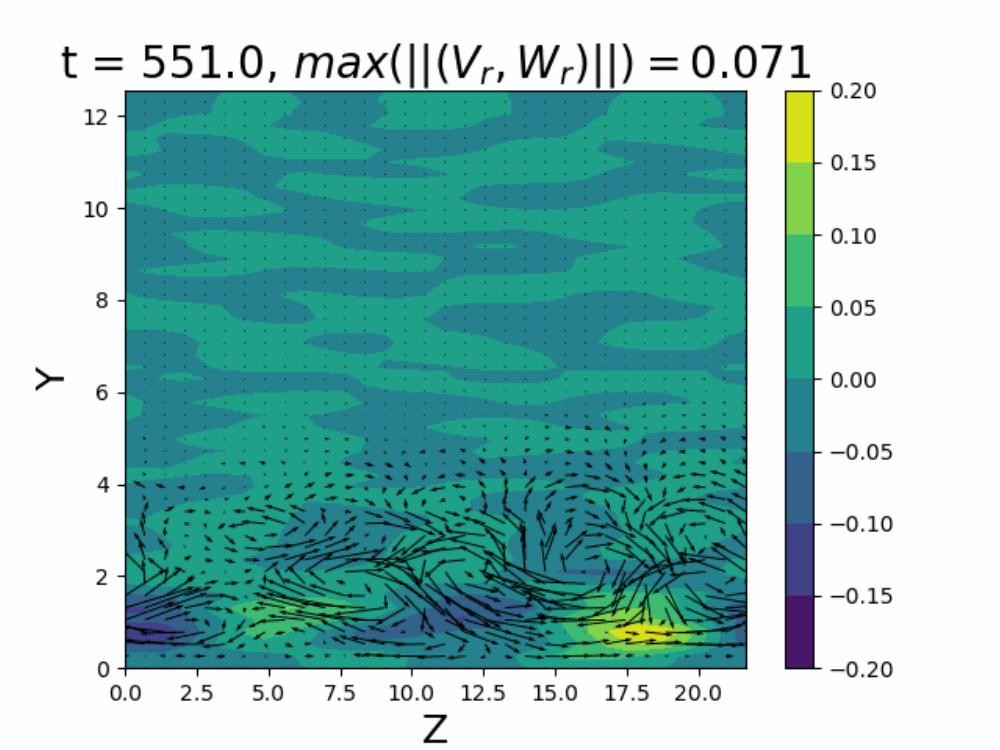}
            \label{subfig:b}%
        }\\
        \subfloat[]{%
            \includegraphics[width=.48\linewidth]{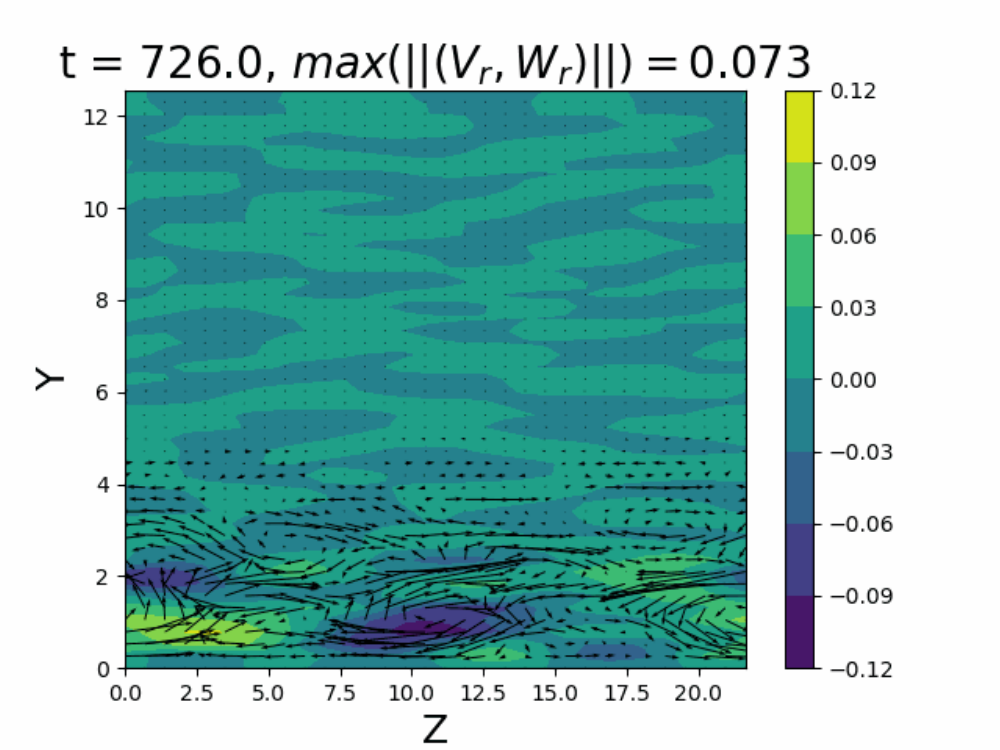}
            \label{subfig:c}%
        }\hfill
        \subfloat[]{%
            \includegraphics[width=.48\linewidth]{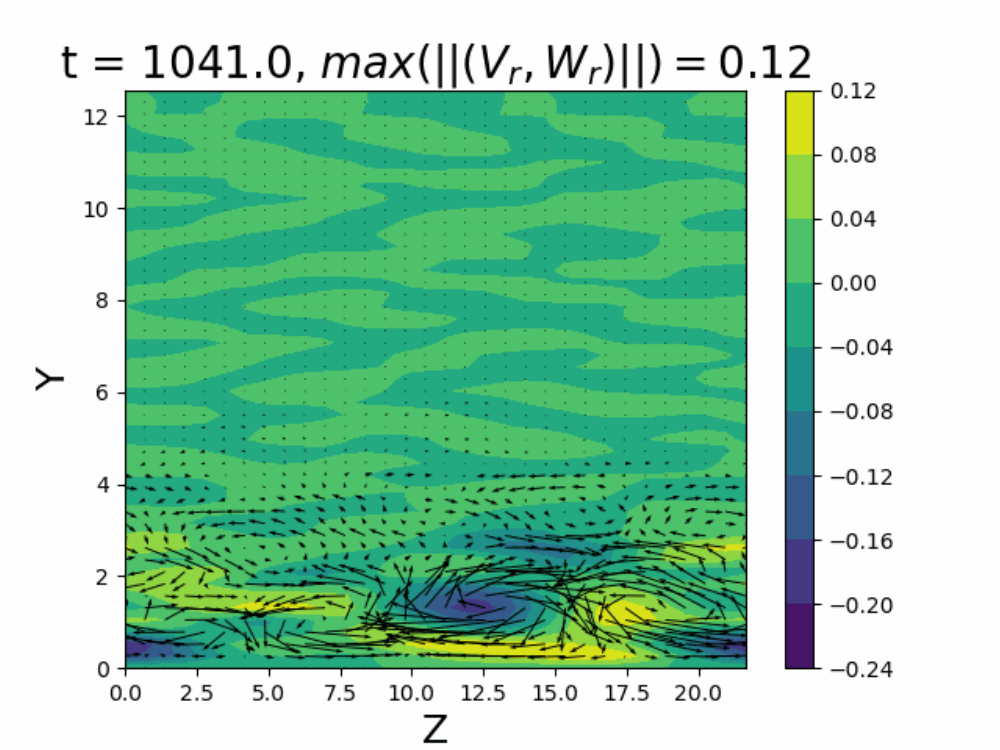}
            \label{subfig:d}%
        }
\caption{Representative snapshots of self sustaining turbulence. Shown in colors are streak and shown in vectors are roll velocity. Parameters: $\epsilon=0$; $Re=300$}
\label{SSPsnapshots}
\end{figure}

\begin{figure}
\centering{
\includegraphics[width=\linewidth]{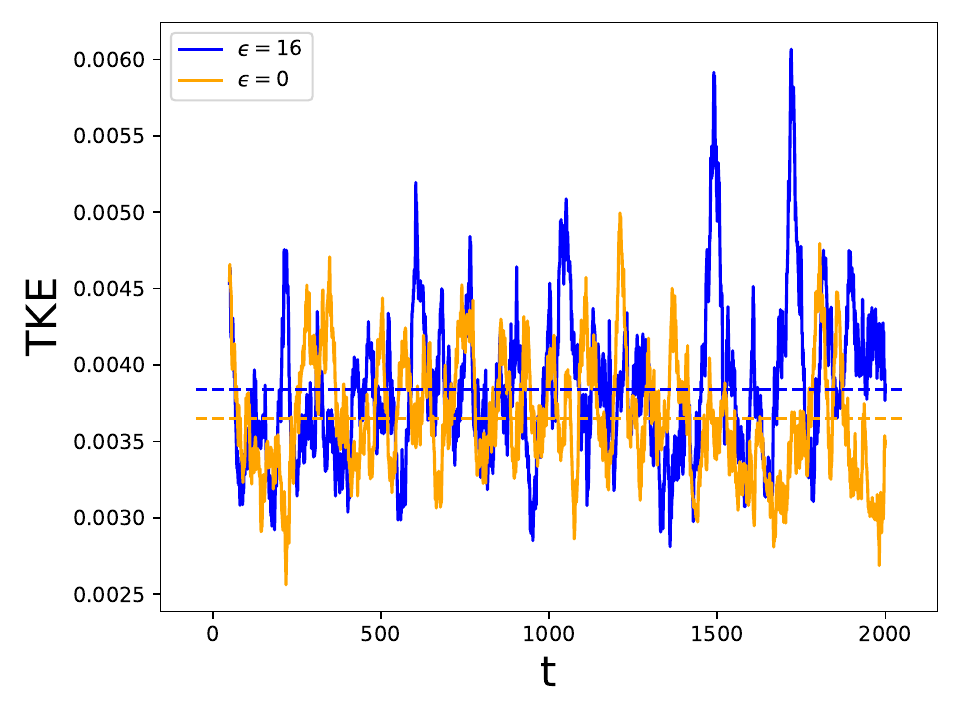}}
\caption{Time series of TKE as a turbulence diagnostic for two cases: turbulence with the background turbulence excitation used to trigger the transition to the turbulent state retained ($\epsilon = 16$, blue), and self-sustained turbulence after the instigating excitation has been removed ($\epsilon = 0$, orange); $Re=300$.}
\label{tkeSSPvsfull}
\end{figure}

We find that all turbulent equilibria in our simulations are self-sustaining. Shown in figure \ref{SSPsnapshots} are representative snapshots of self-sustaining turbulence. 
This result is consistent with our analysis of the maintenance of the RSS and the fluctuation enstrophy, which are indicative of the underlying mechanism sustaining PBL RSS turbulence being the same self-sustaining process (SSP) familiar in the context of the dynamics of wall-bounded shear flow turbulence \citep{Farrell-Ioannou-2012}. 
For example, the S3T SSD PBL RSS dynamics at $\epsilon=16$ equilibrates to a turbulent state. Upon removing the stochastic forcing by setting $\epsilon=0$, this tubulent state persists as verified by comparison between time series of the fluctuation energy, TKE, of these turbulent states shown in figure \ref{tkeSSPvsfull}.  Fluctuation energy is defined as: 
\begin{equation}
TKE(t)=\frac{1}{2}[u'^2+v'^2+w'^2]_{x,y,z}.
\end{equation}

While fluctuation TKE is greater when $\epsilon=16$ than when $\epsilon=0$, both are sustained primarily by the same SSP as that sustaining wall-bounded shear flow turbulence. Greater TKE is expected in the case with stochastic forcing as $\epsilon$ directly injects energy into the perturbation covariance $\bv{C}$

\section{Conclusion}
This study advances understanding of Ekman turbulence in the PBL by identifying RS torque instability as a key mechanism for RSS formation and turbulence maintenance. Employing the S3T SSD framework we recast the turbulent Ekman model as a SSD system. This reformulation provides an analytical unification of the traditional inflectional instability with the newly identified RS torque instability and clarifies their respective roles in turbulent Ekman dynamics.  The inflectional instability is represented in the first cumulant within the S3T SSD,  whereas the RS torque instability emerges from interactions between the first and second cumulants. These two instabilities can therefore be investigated independently.
For instance, we find that RSS turbulence in the PBL can be instigated and supported by the RS torque mechanism alone,  when the inflectional instability has been suppressed.  
A key advantage of the S3T SSD formulation of PBL dynamics is that the nonlinear S3T SSD that allows identification of the instabilities initiating the formation of the RSS structure in a simplified context also contains in the same simplified context the essential mechanism by which these eigenmodes are equilibrated. 
We find that as a function of the background turbulence intensity parameter, $\epsilon$, the S3T SSD RSS equilibrates to a fixed point, a time-dependent state of limited complexity, or to a turbulent state. The turbulent state is found to be self-sustaining in the sense that, once it has been established by bifurcation from an unstable S3T SSD equilibrium, the destabilizing background turbulence excitation parameter, $\epsilon$, can be removed and the turbulence subsequently sustains itself through essentially the same self-sustaining process (SSP) familiar in the context of wall-bounded shear flows.
In the S3T SSD the dynamics of RSS equilibria can be partitioned into physical mechanisms responsible for the statistical mean balance maintaining the isolated streak, roll, and perturbation components. We find in the case of turbulent RSS equilibria that the dominant balance maintaining the streak is growth by the roll-induced lift-up process opposed by a combination of downgradient fluctuation RS and viscous dissipation. The dominant balance maintaining the roll is between RS torque and dissipation. We find Coriolis force term is negligible in balance maintaining both the streak and roll. 
Given the widely observed PBL velocity profiles, which differ substantially from the classical 45° veering predicted by the Ekman spiral, and the inability of inflectional instability alone to account for this discrepancy, we employed our S3T SSD framework to investigate its origin. S3T SSD offers a more comprehensive theoretical description, as it incorporates both the inflectional instability mechanism and the nonlinear S3T SSD Reynolds-stress (RS) torque mechanism.

We find that equilibration driven solely by the inflectional mode alters the Ekman spiral only marginally. In contrast, increasing the influence of the RS torque mechanism by raising $\epsilon$, and ultimately by promoting the PBL into a self-sustaining turbulent state,  produces a further reduction in the veering angle, bringing the surface wind direction into closer alignment with the geostrophic wind and with observations.  In addition, we find improved agreement with the observed collapse of the interior Ekman layer profile as $\epsilon$ is increased.
Importantly, our findings show that the observed PBL velocity profile corresponds to strongly turbulent equilibria in our regime diagram. Consistently, values of turbulence intensity predicted from observations \citep{Friedrich2012, Watkins2012, Duncan2019, Barthelmie2014} also correspond to a strongly turbulent regime. 
Fluxes of momentum and tracers in the PBL are important in a wide range of physical phenomena.  In the case of rapid deepening of a TC an important manifold of RSS instabilities associated with inertial instabilities arises in the inner core.  The methods we applied here to the Ekman problem can be straightforwardly extended to a gradient wind balanced flow and the three way dynamics of the inflectional, inertial and RS associated RSS growth mechanisms examined.  This is a subject of ongoing work.
In summary, an S3T SSD turbulence model that incorporates both the inflectional instability mechanism and the RS torque mechanism  offers a more comprehensive approach for understanding PBL turbulence and improves correspondence with observations.

\section{Appendix}
\subsection{The Individual Components of Equations \eqref{A11eq}, \eqref{A12eq}, \eqref{A21eq}, and \eqref{A22eq}}
\label{appendix:Adetail}
Detailed descriptions of $L_{OS}(U)$, $L_{C1}(U)$, $L_{C2}(U)$, and $L_{SQ}(U)$ can be found in \citet{Farrell-Ioannou-2012}. The remaining components of Eqs.~\eqref{A11eq}, \eqref{A12eq}, \eqref{A21eq}, and \eqref{A22eq} are given by 
\begin{equation}
\begin{aligned}
LV_{11}(V)=(k^2V_y-V_{yzz}+k^2V \partial_y -V_{zz}\partial_y -2V_{yz}\partial_z -V_y \partial_{zz} -2V_z \partial_{yz}-V\partial_{yzz})\\ + [(ik)(V_y \partial_y +V \partial_{yy})](-ik \Delta_2^{-1}\partial_y)\\+(k^2 V_z-V_{zzz}+V_{yz}\partial_y+V_z\partial_{yy}+V_y \partial_{yz}+V \partial_{yyz}-2V_{zz}\partial_z -V_z\partial_{zz})(-\Delta_2^{-1}\partial_{yz})
\end{aligned}
\end{equation}
\begin{equation}
\begin{aligned}
LW_{11}(W)=(k^2W\partial_z+ W_{yy}\partial_z +W_y \partial_{yz}+W_{yyz}+W_{yz}\partial_y-W_{zz}\partial_z-2W_z \partial_{zz}-W\partial_{zzz})\\+(ik W_y \partial_z + ik W \partial_{yz})(-ik \Delta_2^{-1}\partial_y)\\+(2W_{yz}\partial_z+2W_z\partial_{yz}+W_{zzy}+W_{zz}\partial_y+W_{y}\partial_{zz}+W\partial_{zzy})(-\Delta_2^{-1}\partial_{yz})
\end{aligned}
\end{equation}
\begin{equation}
\begin{aligned}
LV_{12}(V) &= (ik)\bigl(V_y \,\partial_y + V \,\partial_{yy}\bigr)\bigl(\Delta_2^{-1}\partial_z\bigr) \\
&\quad - (ik)\bigl(k^2 V_z - V_{zzz} + V_{yz}\partial_y + V_z\partial_{yy} + V_y\partial_{yz} + V\partial_{yyz} 
- 2V_{zz}\partial_z - V_z\partial_{zz}\bigr)\Delta_2^{-1}
\end{aligned}
\end{equation}
\begin{equation}
\begin{aligned}
LW_{12}(W)=(ik W_y \partial_z + ik W \partial_{yz})(\Delta_2^{-1}\partial_z)\\+ (2W_{yz}\partial_z+2W_z\partial_{yz}+W_{zzy}+W_{zz}\partial_y+W_{y}\partial_{zz}+W\partial_{zzy})(-ik \Delta_2^{-1})
\end{aligned}
\end{equation}
\begin{equation}
\begin{aligned}
LV_{21}(V)=-(V_z\partial_y+V\partial_{yz})(-ik\Delta_2^{-1}\partial_y)+(ikV \partial_y)(-\Delta_2^{-1}\partial_{yz})
\end{aligned}
\end{equation}
\begin{equation}
\begin{aligned}
LW_{21}(W)=(ik W_y)-(W_z\partial_z+W\partial_{zz})(-ik\Delta_2^{-1}\partial_y)+(ik)(W_z+W\partial_z)(-\Delta_2^{-1}\partial_{yz})
\end{aligned}
\end{equation}
\begin{equation}
\begin{aligned}
LV_{22}(V)=-(V_z\partial_y+V\partial_{yz})(\Delta_2^{-1}\partial_z)+(ikV \partial_y)(-ik \Delta_2^{-1})
\end{aligned}
\end{equation}
\begin{equation}
LW_{22}(W)=-(W_z\partial_z+W\partial_{zz})(\Delta_2^{-1}\partial_z)+(ik)(W_z+W\partial_z)(-ik \Delta_2^{-1})
\end{equation}
\subsection{Undoing Coordinate Transformation for Equations \eqref{Usurfstresseq} and \eqref{Wsurfstresseq} }
\label{appendix:coordinate}
$[U]_z$ and $[W]_z$ can be written in terms of $[U^{n}]_{z^{n}}$ and $[W^{n}]_{z^{n}}$ as:\\

\begin{equation}
[U]_{z}=cos(\theta) [U^{n}]_{z^{n}}-sin(\theta) [W^{n}]_{z^{n}}
\label{Uzinunrotate}
    \end{equation}
    \begin{equation}
    [W]_{z}=sin(\theta) [U^{n}]_{z^{n}}+cos(\theta) [W^{n}]_{z^{n}}.
    \label{Wzinunrotate}
\end{equation}\\

Using \eqref{Uzinunrotate} and  \eqref{Wzinunrotate} to rewrite equations \eqref{Usurfstresseq} and \eqref{Wsurfstresseq} as:\\

\begin{equation}
\begin{split}
sin(\theta) [U^{n}]_{y,z^{n}}+cos(\theta) [W^{n}]_{y,z^{n}}=-\frac{1}{2\cdot Ly}cos(\theta)\frac{d [U^{n}]_{z^{n}}}{dy}(y=0)+\frac{1}{2\cdot Ly}sin(\theta) \frac{d[W^{n}]_{z^{n}}}{dy}(y=0)\\+sin(\theta)
\label{UsurfeqlongUnWn}
\end{split}
\end{equation}
\begin{equation}
\begin{split}
cos(\theta) [U^{n}]_{y,z^{n}}-sin(\theta) [W^{n}]_{y,z^{n}}=\frac{1}{2\cdot Ly}sin(\theta) \frac{d[U^{n}]_{z^{n}}}{dy}(y=0)+\frac{1}{2\cdot Ly}cos(\theta) \frac{d[W^{n}]_{z^{n}}}{dy}(y=0)\\+cos(\theta).
\label{WsurfeqlongUnWn}
\end{split}
\end{equation}\\

Sum of $cos(\theta)\cdot \eqref{UsurfeqlongUnWn}$ and $-sin(\theta)\cdot \eqref{WsurfeqlongUnWn}$ yields:

\begin{equation}
[W^{n}]_{y,z^{n}}=-\frac{1}{2\cdot Ly}\frac{d[U^{n}]_{z^{n}}}{dy}(y=0)
\end{equation}

\end{document}